\documentclass[preprint,12pt]{elsarticle}

\usepackage{amssymb}
\usepackage{amsmath}
\usepackage{amsthm}

\usepackage{newtx}

\usepackage{comment}
\usepackage{graphicx}
\usepackage{bbold} 
\usepackage{forest}
\usepackage{mathpartir}
\usepackage{stmaryrd}
\usepackage[all]{xy}
\usepackage[hidelinks]{hyperref}
\usepackage[nameinlink,noabbrev,capitalise]{cleveref} 
\usepackage[normalem]{ulem}

\newtheorem{thm}{Theorem}[section]
\newtheorem{prop}[thm]{Proposition}
\newtheorem{corollary}[thm]{Corollary}

{\theoremstyle{definition}
\newtheorem{defn}[thm]{Definition}
}
\crefname{defn}{Definition}{Definitions}

\crefname{appendix}{}{}

\counterwithin{figure}{section}

\journal{J. Log. Algebraic Methods Program.}

\begin{document}

\begin{frontmatter}



\title{Comodule Representations of Second-Order Functionals}


\author[tartu]{Danel Ahman\corref{cor}} 
\cortext[cor]{Corresponding author}
\ead{danel.ahman@ut.ee}

\author[fmf,imfm]{Andrej Bauer} 
\ead{andrej.bauer@andrej.com}

\affiliation[tartu]{organization={Institute of Computer Science, University of Tartu},
            addressline={Narva mnt 18},
            city={Tartu},
            country={Estonia}}

\affiliation[fmf]{organization={Faculty of Mathematics and Physics, University of Ljubljana},
            addressline={Jadranska~19},
            city={Ljubljana},
            country={Slovenia}}

\affiliation[imfm]{organization={Institute of Mathematics, Physics and Mechanics},
            addressline={Jadranska~19},
            city={Ljubljana},
            country={Slovenia}}

\begin{abstract}
We develop and investigate a general theory of representations of second-order functionals,
based on a notion of a right comodule for a monad on the category of containers.
We show how the notion of comodule representability naturally subsumes classic representations of continuous functionals with well-founded trees.
We find other kinds of representations by varying the monad, the comodule, and in some cases the underlying category of containers.
Examples include uniformly continuous or finitely supported functionals, functionals querying their arguments precisely once, or at most once,
functionals interacting with an ambient environment through computational effects, as well as functionals trivially representing themselves.
Many of these rely on our construction of a monad on containers from a monad on shapes and a weak Mendler-style monad algebra on the universe for positions.
We show that comodule representability on the category of propositional containers, which have positions valued in a universe of propositions, is closely related to instance reducibility in constructive mathematics, and through it to Weihrauch reducibility in computability theory.
\end{abstract}




\end{frontmatter}



\newcommand{\ABcomment}[1]{\textcolor{purple}{#1}} 
\newcommand{\DAcomment}[1]{\textcolor{blue}{#1}} 

\newcommand{\defeq}{\mathrel{\overset{\text{\tiny def}}{=}}} 

\makeatletter 
\newcommand{\customlabel}[2]{%
   \protected@write \@auxout {}{\string \newlabel {#1}{{(#2)}{\thepage}{#2}{#1}{}} }%
   \hypertarget{#1}{#2}
}
\makeatother

\newcommand{\prd}[1]{{\textstyle\prod_{#1} \,}} 
\newcommand{\sm}[1]{{\textstyle\sum_{#1} \,}} 
\newcommand{\all}[1]{\forall_{#1}.\,} 
\newcommand{\iall}[1]{\forall_{\{#1\}}.\,} 
\newcommand{\some}[1]{\exists_{\,#1}.\,} 
\newcommand{\lthen}{\Rightarrow} 
\newcommand{\liff}{\Leftrightarrow} 

\newcommand{\iprd}[1]{{\textstyle\prod_{\{#1\}} \,}} 
\newcommand{\ia}[1]{\{#1\}} 

\newcommand{\lam}[1]{\lambda #1.\,} 
\newcommand{\ilam}[1]{\lambda \ia{#1}.\,} 

\newcommand{\of}{{:}} 

\newcommand{\fst}{\mathsf{fst}} 
\newcommand{\snd}{\mathsf{snd}} 
\newcommand{\pair}[2]{\langle #1, #2 \rangle} 

\newcommand{\inl}{\mathsf{inl}} 
\newcommand{\inr}{\mathsf{inr}} 

\newcommand{\inlpat}[1]{\mathsf{inl}(#1)} 
\newcommand{\inrpat}[1]{\mathsf{inr}(#1)} 

\newcommand{\Cont}{\mathbf{Cont}} 
\newcommand{\PCont}{\mathbf{PCont}} 
\newcommand{\Type}{\mathbf{Type}} 
\newcommand{\Prop}{\mathbf{Prop}} 
\newcommand{\U}{\mathbf{U}} 

\newcommand{\Set}{\mathbf{Set}} 

\newcommand{\op}{\mathrm{op}} 

\newcommand{\Coalg}[1]{\mathbf{Coalg}(#1)} 

\newcommand{\Mod}[2]{#1\text{-}\mathbf{Mod}(#2)} 
\newcommand{\CoMod}[2]{#1\text{-}\mathbf{CoMod}(#2)} 

\newcommand{\RFun}[1]{\mathbf{RFun}(#1)} 
\newcommand{\Fun}{\mathbf{Fun}} 
\newcommand{\F}{\mathcal{F}} 

\newcommand{\C}{\mathcal{C}} 
\newcommand{\D}{\mathcal{D}} 
\newcommand{\E}{\mathcal{E}} 

\newcommand{\Cat}{\mathsf{Cat}} 

\newcommand{\obj}[1]{\mathsf{ob}(#1)} 

\newcommand{\One}{\mathbb{1}} 
\newcommand{\Zero}{\mathbb{0}} 
\newcommand{\Two}{\mathbb{2}} 
\newcommand{\Nat}{\mathbb{N}} 

\newcommand{\Fin}{\mathsf{Fin}} 

\newcommand{\finset}[1]{\{\!|#1|\!\}}

\newcommand{\eat}{\mathsf{e}} 
\newcommand{\cook}{\text{c}} 
\newcommand{\tree}{\mathsf{t}} 

\newcommand{\alg}{\alpha} 

\newcommand{\shapeMap}[1]{#1_\mathsf{shp}} 
\newcommand{\positionMap}[1]{#1_\mathsf{pos}} 

\newcommand{\shapeMapSup}[1]{#1^\mathsf{shp}} 
\newcommand{\positionMapSup}[1]{#1^\mathsf{pos}} 

\newcommand{\tinyp}{\text{\textnormal{\tiny p}}}
\newcommand{\tinyc}{\text{\textnormal{\tiny c}}}
\newcommand{\tinyr}{\text{\textnormal{\tiny r}}}
\newcommand{\tinyw}{\text{\textnormal{\tiny w}}}

\newcommand{\cont}[2]{#1 \mathbin{\lhd} #2} 
\newcommand{\pcontsymbol}{\mathbin{\lhd^\tinyp}} 
\newcommand{\pcont}[2]{#1 \pcontsymbol#2} 
\newcommand{\dcont}[2]{#1 \mathbin{\lhd_{D}} #2} 


\newcommand{\sem}[1]{\llbracket #1 \rrbracket} 
\newcommand{\cosem}[1]{\langle\!\langle #1 \rangle\!\rangle} 

\newcommand{\id}{\mathsf{id}} 
\newcommand{\Id}{\mathsf{Id}} 

\newcommand{\comp}{\mathbin{\circ}} 

\newcommand{\Tree}[1]{\mathsf{Tree}(#1)} 
\newcommand{\leaf}{\mathsf{leaf}} 
\newcommand{\node}{\mathsf{node}} 
\newcommand{\graft}{\mathsf{graft}} 
\newcommand{\trs}[1]{#1} 

\newcommand{\Path}[2][]{\mathsf{Path}_{#1}(#2)} 
\newcommand{\PathFam}[1][]{\mathsf{Path}_{#1}} 
\newcommand{\pstop}{\mathsf{stop}} 
\newcommand{\pstep}{\mathsf{step}} 
\newcommand{\pfst}{\mathsf{pfst}} 
\newcommand{\psnd}{\mathsf{psnd}} 
\newcommand{\pth}[1]{\vec{#1}} 

\newcommand{\Tr}{\mathcal{T}} 

\newcommand{\Pow}[1]{\mathcal{P}(#1)} 
\newcommand{\FPow}[1]{\mathcal{P}_{\mathsf{f}}#1} 
\newcommand{\IPow}[1]{\mathcal{P}_{\!+}#1} 

\newcommand{\IO}[1]{\mathsf{IO}#1} 
\newcommand{\inp}{\mathsf{inp}} 
\newcommand{\coinp}[1]{\mathsf{inp}_{#1}} 
\newcommand{\out}{\mathsf{out}} 
\newcommand{\coout}[1]{\mathsf{out}_{#1}} 
\newcommand{\return}{\mathsf{ret}} 

\newcommand{\IOTrace}[2][]{\mathsf{Trace}_{#1}(#2)} 
\newcommand{\iostop}{\mathsf{stop}} 
\newcommand{\istep}{\mathsf{istep}} 
\newcommand{\ostep}{\mathsf{ostep}} 
\newcommand{\trc}{\tau} 

\newcommand{\IOTree}[1]{\mathsf{IOTree}(#1)} 
\newcommand{\IOPath}[2][]{\mathsf{IOPath}_{#1}(#2)} 

\newcommand{\St}{\mathsf{St}} 
\newcommand{\Rd}{\mathsf{Rd}} 
\newcommand{\Exc}{\mathsf{Exc}} 
\newcommand{\ExcT}{\mathsf{ExcT}} 

\newcommand{\Trns}{\mathsf{Tr}} 

\newcommand{\Delay}{\mathsf{D}} 
\newcommand{\dnext}{\mathsf{next}} 
\newcommand{\dnow}{\mathsf{now}} 

\newcommand{\R}{\mathsf{R}} 

\newcommand{\cId}{\mathsf{Id}^\tinyc} 
\newcommand{\cComp}[2]{#1 \mathbin{\circ^\tinyc} #2} 

\newcommand{\cOne}{\mathbb{1}^\tinyc} 
\newcommand{\cZero}{\mathbb{0}^\tinyc} 

\newcommand{\cProd}[2]{#1 \mathbin{\times^\tinyc} #2} 
\newcommand{\cSum}[2]{#1 \mathbin{+^\tinyc} #2} 

\newcommand{\cExp}[2]{#1 \mathbin{\Rightarrow^\tinyc} #2} 
\newcommand{\wev}{\mathsf{ev}} 

\newcommand{\wcurry}[1]{\hat{#1}} 

\newcommand{\pOne}{\mathbb{1}^\tinyp} 
\newcommand{\pZero}{\mathbb{0}^\tinyp} 
\newcommand{\wExp}[2]{#1 \mathbin{\Rightarrow^\tinyw} #2} 
\newcommand{\pExp}[2]{#1 \mathbin{\Rightarrow^\tinyp} #2} 
\newcommand{\pProd}[2]{#1 \mathbin{\times^\tinyp} #2} 
\newcommand{\pSum}[2]{#1 \mathbin{+^\tinyp} #2} 

\newcommand{\pId}{\mathsf{Id}^\tinyp} 

\newcommand{\cInl}{\mathsf{inl}^\tinyc} 
\newcommand{\cInr}{\mathsf{inr}^\tinyc} 
\newcommand{\cCopair}[2]{[#1,#2]^\tinyc} 

\newcommand{\rProd}[2]{#1 \mathbin{\times^\tinyr} #2} 
\newcommand{\rFst}{\mathsf{fst}^\tinyr} 
\newcommand{\rSnd}{\mathsf{snd}^\tinyr} 
\newcommand{\rPair}[2]{\langle #1 , #2 \rangle^\tinyr} 

\newcommand{\bind}[1]{#1^\dagger} 

\newcommand{\malg}[1]{#1^\star} 
\newcommand{\palg}[1]{#1^{+}} 

\newcommand{\malgfun}[1]{[#1]^\star} 
\renewcommand{\i}{\mathsf{i}} 
\renewcommand{\j}{\mathsf{j}} 
\renewcommand{\k}{\mathsf{k}} 

\newcommand{\malgIO}[1]{\overline{#1^{\star}}} 

\newcommand{\cosemIO}[2]{\cosem{#1}_{#2}} 

\newcommand{\ileq}{\leq_{\mathsf{I}}} 



\forestset{
  tria/.style={
    node format={
      \noexpand\node [
      draw,
      shape=regular polygon,
      regular polygon sides=3,
      rotate=180,
      inner sep=0pt,
      outer sep=0pt,
      \forestoption{node options},
      anchor=corner 1
      ]
      (\forestoption{name}) {\rotatebox{180}{\foresteoption{content format}}};
    },
    child anchor=corner 1,
  },
}%



\definecolor{light-gray}{gray}{0.85}

\newcommand{\dcomment}[1]{{
  \hypersetup{linkcolor=black}
  \colorbox{light-gray}{{\fontsize{0.25cm}{0.25cm}$\text{#1}$}}}}



\section{Introduction}
\label{sec:introduction}

A \emph{second-order functional} is a mapping that takes functions as arguments. Examples in mathematics and computer science are plentiful: integration, differentiation, supremum and infimum, summation, limit, logical quantifiers, maps and folds from functional programming, stream transformers, signal filters, etc.

In many cases of interest, such functionals have additional properties that allow us to \emph{represent} them in useful ways.

The prototypical example is a continuous functional $F : \Nat^\Nat \to \Nat$ from the
Baire space $\Nat^\Nat$ to natural numbers~$\Nat$, equipped with the product and discrete topologies, respectively.
As was already known to Brouwer~\cite{brouwer27:_uber_defin_funkt}, such a functional may be represented by a well-founded countably-branching \emph{question-answer} tree~$\tree_F$, whose nodes, branches, and leaves are labelled with numbers. To compute~$F(\alpha)$ at $\alpha : \Nat^\Nat$, we start at the root of~$\tree_F$ and climb the tree by taking the $\alpha(k)$-th branch if the current node is labelled with~$k$. As the tree is well-founded, we will eventually encounter a leaf labelled with an~$n$, at which point we know that $F(\alpha) = n$.

A second example of representable second-order functionals arises in logic. A proof of an implication between universally quantified statements
\begin{equation*}
  (\all{a \of A} \phi\,a) \lthen (\all{b \of B} \psi\,b)
\end{equation*}
may be quite arbitrary, but in practice one mostly encounters proofs by a \emph{functional instance reduction},
namely a map $f : B \to A$ is given such that $\phi(f\,b) \lthen \psi\,b$ holds for all $b : B$.
The map~$f$ together with a proof of $\all{b \of B} \phi(f\,b) \lthen \psi\,b$, may be construed as a representation for the implication.
For instance, the usual proof showing that Zorn's lemma implies the Axiom of choice proceeds precisely in this fashion: given a family of inhabited sets, we construct an instance of a chain-complete poset, namely the partial choice maps for the given family, apply Zorn's lemma to obtain a maximal element, and argue that it is a total choice map.

We analyse and organise such phenomena into a general theory of representations of second-order functionals,
based on a notion of a right comodule for a monad on the category of containers.
The scene is set in \cref{sec:type-theor-prel}, where we give a brief account of the type-theoretic setup that we adopt for the rest of the paper.
In \cref{sec:tree-representation}, we recast the example of continuous functionals and well-founded trees to make them amenable to a category-theoretic formulation.
After reviewing containers, monads and comodules in \cref{sec:containers-monads-comodules}, we use them in \cref{sec:comodule-representation-of-functionals} to formulate our central notion -- comodule representations of second-order functionals. We show that such representations themselves are the objects of a Kleisli category for a monad that maps fully onto the category of represented functionals.
In \cref{sec:examples}, we revisit the motivating example, and find several new ones.
\Cref{sec:effectful-functionals} studies representations of functionals in the presence of computational effects.
\Cref{sec:propositional-containers} develops a variation of functional instance
reductions that pertains to instance reducibility
\cite{Bauer:InstanceReducibility} in constructive mathematics, and through it to
Weihrauch reducibility~\cite{brattka11:_weihr} in computability theory. We
review related work in \cref{sec:related-work} and conclude in
\cref{sec:conclusion}. In \cref{appendix:definition-of-tree-monad}, we give a
detailed definition of the well-founded tree monad on containers for
representing continuous functionals.

\paragraph*{Acknowledgments}
We are thankful to Richard Garner and Mauro Jaskelioff for interesting
discussions about our and their work on representations of second-order
functionals. We also thank the anonymous reviewers for their hard work and
interesting suggestions.
We thank Richard Garner and a reviewer for suggesting an equivalent formulation of
comodule representations in terms of representable functors and monad algebras.
This work received support from the COST Action EuroProofNet
(CA20111).
This work was supported by the Estonian Research Council grant PRG2764.
This material is based upon work supported by the Air Force Office of Scientific Research under award number FA9550-21-1-0024.

\section{Type-Theoretic Preliminaries}
\label{sec:type-theor-prel}

Our development can be understood equally well as taking place in a type-theoretic or a set-theoretic foundation.
However, the topic lends itself quite naturally to a type-theoretic mode of speech, which we duly adopt.
While we do not wish to be too specific about an underlying formalism, for the sake of concreteness we posit it to be  Martin-Löf type theory~\cite{MartinLoef:IntuitionisticTypeTheory,MartinLoef:IntuitionsiticTheoryOfTypes} with dependent products $\prd{x \of A} B(x)$, dependent sums $\sm{x \of A} B(x)$, the identity types, written just as equality $a = b$, and any types a given situation demands, such as the natural numbers~$\Nat$, a type universe~$\Type$, or an impredicative universe of propositions~$\Prop$.

To keep the exposition to the point, we thwart type-theoretic idiosyncrasies pertaining to the identity types by working in an extensional type theory that enjoys whatever extensionality principles accomplish the job.

To aid readability, we use the Agda-style notation for implicit arguments: an
argument whose declaration is surrounded by curly braces is \emph{implicit} in
the sense that we need not specify it in an application, as it can be inferred
from the types of other arguments. For example, function composition is defined as
\begin{equation*}
  {-} \comp {-} \defeq \ilam{A, B, C \of \Type} \lam{(g \of B \to C) \, (f \of A \to B) \, (x \of A)} g (f \, x).
\end{equation*}
The curly braces indicate that the type arguments $A$, $B$, $C$ are elided,
so that we write $g \comp f$ instead of $g \comp_{A,B,C} f$. The implicit arguments also appear in its type:
\begin{equation*}
  {-} \comp {-} : \prd{\ia{A,B,C \of \Type}} (B \to C) \to (A \to B) \to A \to C.
\end{equation*}


\section{Tree Representation of Second-Order Functionals}
\label{sec:tree-representation}

As a first step, we generalise the simply-typed functionals $F : \Nat^\Nat \to \Nat$ from
\cref{sec:introduction} to ones operating on general dependent products, and use them both as domains and codomains, so that functionals can be composed.
As such, the general form of a \emph{second-order functional} under consideration in this paper is
\[
  F : \left(\prd{a \of A} P\, a \right) \to \left(\prd{b \of B} Q\, b \right),
\]
where the dependent products range over type families $P : A \to \Type$ and $ Q: B
\to \Type$. This matches a recent type-theoretic study of continuous
functionals~\cite{baillon22:_garden_pythia_model_contin_depen_settin,baillon23:_contin_type_theor}.

Next, we review constructions of inductively defined trees and the associated notion of paths.
Given a type $A : \Type$ and a type family $P : A \to \Type$, let the type $\Tree{A,P}$ of \emph{$(A,P)$-trees} be generated inductively by the clauses
\begin{mathpar}
  \inferrule 
    {\phantom{\Tree{A,P}}} 
    {\leaf : \Tree{A,P}}
  
  \inferrule 
    {a : A \qquad \trs{t} : P\, a \to \Tree{A,P}} 
    {\node(a,\trs{t}) : \Tree{A,P}}
\end{mathpar}
That is, an $(A,P)$-tree is either an unlabelled leaf, or a tree $\node(a, \trs{t})$ with root labelled by~$a$ and $(P \, a)$-many subtrees~$\trs{t}$.
Such trees arise naturally as \emph{$\mathcal{W}$-types}~\cite{Martin-Loef:ConstructivemathsAndComputerProgramming,Moerdijk:WellFoundedTrees} for certain containers, a direction we elaborate on in \cref{sec:tree-monad}.

The reader may have noticed that leaves are not labelled, contrary to the trees used in \cref{sec:introduction}.
We shall keep the leaf labelling separate from the tree.
In order to gain access to the leaves, we associate with any tree $t : \Tree{A,P}$ the \emph{paths} $\Path[A,P]{t}$ from the root of $t$ to its leaves, defined inductively by the clauses
\begin{mathpar}
  \inferrule
     {\phantom{\Path[A,P]{t}}} 
     {\pstop : \Path[A,P]{\leaf}}
    
  \inferrule 
     {p : P\, a \qquad \pth{p} : \Path[A,P]{\trs{t}\, p}} 
     {\pstep(p,\pth{p}) : \Path[A,P]{\node(a,\trs{t})}}
\end{mathpar}
A path has either arrived at a leaf, or it steps along one of the branches~$p$ at the root, and proceeds inductively along a path $\pth{p}$ towards a leaf in the subtree $\trs{t}\,p$.
(The vector arrow in $\pth{p}$ is just a notational suggestion that a path can be thought of as a sequence of steps through the branches of the tree.)
Note that paths always lead all the way to leaves, they cannot terminate at a non-leaf node, and that there is a unique path leading to each leaf.
We occasionally take the liberty of identifying a leaf with the path leading to it.

Given $h : \prd{a \of A} P\, a$ and $t : \Tree{A,P}$, we may recursively compute a path $\cook_{A,P} \, h \, t : \Path[A,P]{t}$ by selecting branches according to~$h$:
\[
\begin{array}{l c l}
  \cook_{A,P}\, h\, \leaf & \defeq & \pstop,
  \\[3pt]
  \cook_{A,P}\, h\, (\node(a,t)) & \defeq &
  \pstep\big(h\, a, \cook_{A,P}\, h\, (t\, (h\, a))\big).
\end{array}
\]
The above defines a map
\[
  \cook_{A,P} : \left(\prd{a \of A} P\, a\right) \to \prd{t \of \Tree{A,P}} \Path[A,P]{t}
\]
that converts its argument~$h$ to a path-forming map $\cook_{A,P}\, h$.

We now have sufficient arboreal tools to proceed with representations.
A \emph{tree representation} of a functional
\begin{equation*}
  F : \left(\prd{a \of A} P\, a \right) \to \left(\prd{b \of B} Q\, b\right)
\end{equation*}
consists of maps (notice that in $\eat_F$ the argument $b$ is implicit)
\begin{equation*}
\tree_F : B \to \Tree{A,P}
\qquad\text{and}\qquad
\eat_F : \iprd{b \of B} (\Path[A,P]{\tree_F\, b} \to Q\, b), 
\end{equation*}
such that the following diagram commutes:
\begin{equation}
  \label{dia:tree-representation}
  \begin{gathered}
  \xymatrix@C=1em@R=3em@M=0.5em@L=0.5em{
    {\prd{a \of A} P\, a}
    \ar[rr]^-{F}
    \ar@/_1pc/[dr]_<<<<<<{\cook_{A,P}}
    &&
    {\prd{b \of B} Q\, b}
    \\
    &
    {\prd{t \of \Tree{A,P}} \Path[A,P]{t}}
    \ar@/_1pc/[ur]_>>>>>{\lam{\alpha} \lam{b} \eat_F\, (\alpha\, (\tree_F\, b))}
  }
  \end{gathered}
\end{equation}
In equational form, the diagram tells us how to compute values of~$F$:
\[
  F\, h\, b = \eat_F\, (\cook_{A,P}\, h\, (\tree_F\, b)).
\]
That is, to compute $F \, h$ at~$b$, we first use $\cook_{A,P}\, h$ on  $\tree_F\, b$ to  select a path (equivalently, a leaf) and then use the value that~$\eat_F$ associates with it. The tree representation of~$F$ is not unique, as one can modify a tree with additional (useless) branching.

Such representations differ from the traditional tree representation of a functional $\Nat^\Nat \to \Nat$ in two respects.

First, $F$ is represented by a \emph{family} of trees, one for each $b : B$.
We may fall back to a single tree by setting~$B$ to the unit type~$\One$, which also
reduces the codomain to a simple type~$Q$, i.e., we would consider functionals $F : \left(\prd{a \of A} P\, a\right) \to Q$.
It should also be noted that some authors~\cite[\S4.6]{troelstra88:_const} represent a functional $F : \Nat^\Nat \to \Nat^\Nat$ by a single tree $\node(0, \tree_F)$ made of the representing tree family $\tree_F : \Nat \to \Tree{\Nat, \lam{n} \Nat}$, which is however possible only when $A = B$ and $P = Q$.

The second difference is that we decompose the representations into a family of trees~$\tree_F$ with unlabelled leaves and a separate leaf labeling map $\eat_F$. The purpose of such a bookkeeping manoeuvre shall become clear in \cref{sec:comodule-representation-of-functionals}, where~$\tree_F$ and~$\eat_F$ turn out to be the components of a container morphism.


\section{Containers, Monads, and Comodules}
\label{sec:containers-monads-comodules}

Before recasting the tree representations of second-order
functionals abstractly, we review the relevant ingredients.
For convenience, and to indicate how our work can be accommodated in type theory, we construe the
universe $\Type$ as a category whose objects are (closed) types $A$ and morphisms are maps
$f : A \to B$. To secure the soundness of such an approach, we work in extensional type type theory.
In any case, we do not wish to dwell on foundational issues, and leave a meta-theoretic
analysis of the setup needed to carry out all the constructions for another occasion.

\subsection{Containers}
\label{sec:containers}

Containers~\citep{AbbottAG:Containers}, also known as
 (univariate) polynomials~\citep{GambinoK:PolyMonads,PolyBook}, were originally used to
represent and work with parametric data types and polymorphic functions between
them. Their scope has expanded to other areas of mathematics and computer science,
ranging from generalised algebraic
signatures~\cite{Martin-Loef:ConstructivemathsAndComputerProgramming},
descriptions of view-update problems in database theory~\cite{AhmanU:UpdateLenses,OConnor:CoalgebraicLenses,Power:CoalgebraicArrays},
to applications in machine learning~\citep{Cruttwell2022:CatThInML,Spivak:LearnersLanguages}. In this paper, we shall employ them
to build general representations of second-order functionals.

\subsubsection{Containers and Container Morphisms}
\label{sec:containers-morphisms}

A \emph{container} $\cont A P$ comprises a type $A$ of \emph{shapes} and a
family $P : A \to \Type$ of \emph{positions}. The terminology stems from
representations of parametric data types: the elements of $A$ denote the
shapes of data structures (e.g., the length of a list, or the structure of a binary tree),
and the positions denote the placeholders within the structure where data resides
(e.g., the indices of a list, the nodes of a tree, or just the leaves of a tree). For example, lists are represented
by $\mathsf{List}^\mathrm{c} \defeq \cont {\mathbb{N}} {\lam{n} \finset{0, 1, \ldots, n }}$
(here and elsewhere we superscript with~${}^\textrm{c}$ notions pertaining to containers, and we write finite sets as $\finset{x_1, \ldots, x_n}$ to avoid confusion with the Agda-style $\ia{-}$ notation for implicit arguments).
Many variations of containers are known, in which shapes or positions may be equipped with additional structure
or be subject to further restrictions. We shall use some such variations in \cref{sec:propositional-containers}.

A \emph{container morphism} $\cont f g : \cont A P \to \cont B Q$ is given by a
\emph{shape map} $f : A \to B$ and a \emph{position map} $g : \iprd{a \of A} (Q\, (f\, b) \to
P\, a)$. The shape map is oriented in the forward direction, but the position map is reversed, because in their original use as representing polymorphic maps between data types, morphisms fill positions in the codomain with data copied from positions in the domain.
The same idea motivates the definition of the composition of container morphisms. Given container morphisms $\cont f g : \cont A P \to \cont B Q$ and $\cont k l : \cont B Q \to \cont C R$, their
\emph{composition} $(\cont k l) \comp (\cont f g) : \cont A P \to \cont C R$ is given by composing the shape and position maps, each in their respective direction:
\[ 
  (\cont k l) \comp (\cont f g) 
  \defeq 
  \cont {(k \comp f)} {\big(\ilam{a} (g\, \ia{a}) \comp (l\, \ia{f\, a})\big)}.
\]
One can easily check that composition is associative, and that the
\emph{identity} morphisms are pairs of identities ${\cont {\id_A} {(\ilam{a} \id_{P\, a})} : \cont A P \to \cont A P}$.
Consequently, containers and container morphisms form a category $\Cont$, which turns
out to have a rich structure. 

The \emph{terminal} and \emph{initial objects} are given by $\cOne \defeq \cont
\One {\lam{\_} \Zero}$ and $\cZero \defeq \cont \Zero {\lam{\_} \Zero}$,
respectively.
Binary \emph{products} and \emph{coproducts} are given by:
\begin{align}
  \label{eq:cProd}
  \cProd {(\cont A P)} {(\cont B Q)} &\;\defeq\;
    \cont {(A \times B)} {\big(\lam{(a,b)} P\, a + Q\, b\big)}, \\
  \cSum {(\cont A P)} {(\cont B Q)} &\;\defeq\;
    \cont {(A + B)} {[P , Q]}, \notag
\end{align}
where $[P , Q] : A + B \to \Type$ is the \emph{copairing} of $P : A \to \Type$
and $Q : B \to \Type$, characterised by $[P,Q]\,(\inl\, a) = P\,a$ and $[P,Q]\,(\inr\, b) =
Q\,b$.
Moreover, it turns out that $\Cont$ has all \emph{small limits} and \emph{colimits}, see~\cite{PolyBook} for details.

The category $\Cont$ is also cartesian-closed, with \emph{exponentials}
\begin{equation}
  \label{eq:cont-exponential}%
  \cExp {(\cont A P)} {(\cont B Q)}
  \; \defeq\; 
  \begin{aligned}[t]
    &\cont
    {\big(\prd{a \of A} \sm{b \of B} (Q\, b \to \One + P\, a) \big)}
    {{}} \\[1pt]
    &{\big(\lam{f} \sm{a \of A} \sm{q \of Q\, (\fst\, (f\, a))} 
    \snd\, (f\, a)\, q = \inl \star\big)}.
  \end{aligned}
\end{equation}
The shape of the exponential encodes the type of container
morphisms from $\cont A P$ to $\cont B Q$, except that summing with~$\One$ is
used to mark positions that remain unfilled. These positions are used in the adjunction
\begin{equation*}
  \mprset{fraction={===}}
  \inferrule 
    {\cProd{(\cont C R)}{(\cont A P)} \to \cont B Q} 
    {\cont C R \to \cExp {(\cont A P)} {(\cont B Q)}}
\end{equation*}
to store data from the container $\cont C R$, see~\citep{AltenkirchLS:HigherOrderContainers}.

Apart from the cartesian and cocartesian monoidal structures, $\Cont$ is also equipped with
the monoidal structure based on \emph{composition of containers}:
\[
  \cComp {(\cont A P)} {(\cont B Q)} \,\defeq\, 
    \cont 
      {\big(\sm{a \of A} (P\, a \to B)\big)} 
      {\big(\lam{(a,v)} \sm{p \of P\, a} Q\, (v\, p)\big)}.
\]
Intuitively, the composed container $\cComp {(\cont A P)} {(\cont B Q)}$
represents a data type of data types, whose outer structure is
described by $\cont A P$ and the inner one by $\cont B Q$.
This monoidal structure is not commutative, but it has a unit, namely the \emph{identity container}
$\cId \defeq \cont {\One} {\lam{\_} \One}$. We refer the reader to \cite{PolyBook} for details on this and other topics pertaining to~$\Cont$.

\subsubsection{Interpretation and Cointerpretation of Containers}
\label{sec:interp-cointerp}

The interpretation of a container $\cont A P$ as a data type is embodied by the
\emph{polynomial endofunctor} $\sem{\cont A P} : \Type \to \Type$, given by
\[
\begin{array}{l c l}
  \sem{\cont A P}\, X & \defeq & \sm{a \of A} (P\, a \to X),  \\[3pt]
  \sem{\cont A P}\, h \, (a,v) & \defeq & (a, h \comp v).
\end{array}
\]
We think of $\sem{\cont A P}\, X$ as the data type storing values of type~$X$ at
the positions of the container, i.e., its elements are pairs $(a, v)$ where $a$
is a shape and $v$ the assignment of values in $X$ to positions in $P\,a$.
For instance, the aforementioned list container yields
$
    \sem{\mathsf{List}^\mathrm{c}}\, X
  = \sm{n \of \Nat} (\Fin\, n \to X)
$,
which is naturally isomorphic to the usual inductive definition
$\mathsf{List}\, X \defeq \mu Z .\, \One + X \times Z$. 
(The notation $\mu Z .\, F(Z)$ stands for the carrier of the initial algebra of the endofunctor $F$.)

The functorial action of the interpretation maps a container
morphism $\cont f g : \cont A P \to \cont B Q$ to a natural transformation
$\sem{\cont f g} : \sem{\cont A P} \to \sem{\cont B Q}$, whose component at~$X$
is $\sem{\cont f g}_X\, (a,v) \defeq (f\, a , \lam{q} v\, (g\, q))$.
Thus $\cont f g$ models what we intuitively take to be a polymorphic function
between parametric data types.

In this way, the \emph{interpretation of containers} itself forms a functor
$\sem{-} : \Cont \to [\Type,\Type]$, from $\Cont$ to the
category of endofunctors on $\Type$. We refer the reader to
\citep{PolyBook,AbbottAG:CategoriesOfContainers} for a study of~$\sem{-}$, such as the preservation of
products $\cProd{}{}$, coproducts $\cSum{}{}$, and composition~$\cComp{}{}$.

For our purposes, the less-studied dual concept will be much more relevant.
Following Ahman and Uustalu's work on (dependently typed) update
monads~\citep{AhmanU:UpdateMonads}, we define the \emph{cointerpretation of
containers} to be given by the contravariant functor $\cosem{-} :
\Cont^{\op} \to [\Type,\Type]$,  whose action on objects is defined as
\[
\begin{array}{l c l}
  \cosem{\cont A P}\, X & \defeq & \prd{a \of A} (P\, a \times X),  \\[3pt]
  \cosem{\cont A P}\, k \, l & \defeq & \lam{(a \of A)} (\fst\, (l \, a), k \, (\snd \, (l\, a))), 
\end{array}
\]
where $k : X \to Y$.
Its action on a container morphism $\cont f g : \cont A P \to \cont B Q$ is
\begin{equation*}
  \cosem{\cont f g}_X\, h \defeq \lam{(a \of A)} \big(g\, (\fst\, (h\, (f\, a)))
, \snd\, (h\, (f\, a))\big).
\end{equation*}

Cointerpretation~$\cosem{-}$ takes the initial $\cZero$ and terminal $\cOne$ containers to the terminal and initial functors, respectively,
\begin{align*}
 \cosem{\cZero}\, X &= \prd{z \of \Zero} (\Zero \times X) \cong \One,\\
 \cosem{\cOne}\, X &= \prd{z \of \One} (\Zero \times X) \cong \prd{z \of \One} \Zero \cong \Zero,
\end{align*}
and the coproduct of containers to the product of functors,
\begin{align*}
    \cosem{\cSum {(\cont A P)} {(\cont B Q)}}\, X
  &= \cosem{\cont {(A + B)} {[P , Q]}}\, X \\
  &= \prd{z : A + B} \big([ P , Q ]\, z \times X\big) \\
  &\cong \big(\prd{a \of A} (P\, a \times X)\big) \times \big(\prd{b \of B} (Q\, b \times X)\big) \\
  &= \cosem{\cont A P}\, X \times \cosem{\cont B Q}\, X.
\end{align*}
Unfortunately, good news ends here. Unlike the interpretation~$\sem{-}$, the cointerpretation is~$\cosem{-}$ is neither full nor faithful, and it only preserves products laxly:
\begin{align*}
    \cosem{\cont A P}\, X + \cosem{\cont B Q}\, X
  &= (\prd{a \of A} P\, a \times X) + (\prd{b \of B} Q\, b \times X) \\
  &\to \prd{(a,b) \of A \times B} (P\, a \times X + Q\, b \times X) \\
  &\cong \prd{(a,b) \of A \times B} ((P\, a + Q\, b) \times X) \\
  &= \cosem{\cont {A \times B} {\lam{(a,b)} P\, a + Q\, b}}\, X \\
  &= \cosem{\cProd {(\cont A P)} {(\cont B Q)}}\, X, 
\end{align*}
where the map in the second row is $[h \mapsto \inl \comp h \comp \fst, g \mapsto \inr \comp g \comp \snd]$.

\subsection{Monads and Comodules}

We review the remaining category-theoretic ingredients that we shall need: monads~\citep{Manes:AlgTheories,Barr:Toposes}, their Kleisli categories, and their (co)modules.

We shall primarily work with a monad on a category~$\C$ given in the form of a
\emph{Kleisli triple} $(T,\eta,\bind{(-)})$, sometimes also called a monad in a
\emph{no-iteration form}~\cite{Marmolejo:NoIterationMonads}, where $T : \obj \C
\to \obj \C$ is a map on objects, $\eta$ a family of maps $\eta_A : A \to T
A$, one for each~$A$ in $\C$, and the \emph{Kleisli extension~$\bind{(-)}$}
takes any map $f : A \to T B$ to a map $\bind f : T A \to T B$. These must satisfy the following laws:
\begin{align*}
  \bind {\eta_A} &= \id_{T A}, &
  \bind f \comp \eta_A &= f, &
  \bind {(\bind g \comp f)} &= \bind g \comp \bind f,
\end{align*}
where $f : A \to TB$ and $g : B \to T C$.
The first and second law say that $\eta$ is the unit for $\bind{(-)}$, and the third that $\bind{(-)}$ is associative.
It follows that~$T$ is the object part of an endofunctor $T : \C \to \C$ and that~$\eta$ is
natural. When the monad structure is clear from the context, we refer to $(T,\eta,\bind{(-)})$ simply by~$T$.

The formulation of monads in terms of Kleisli triples is certainly familiar to functional programmers.
At times we shall also use an equivalent formulation, which is standard in mathematical circles.
That is, a \emph{monad} $(T,\eta,\mu)$ on a category~$\C$ is just a monoid in the
category of endofunctors $[\C,\C]$. Explicitly, it is given by an endofunctor $T : \C \to \C$, and
natural transformations $\eta : \Id \to T$ (the \emph{unit}) and $\mu : T \comp
T \to T$ (the \emph{multiplication}), such that the following diagrams commute:
\[
\xymatrix@C=3em@R=3em@M=0.5em{
T 
  \ar[r]^-{\eta T} 
  \ar@{=}[dr]
& 
T \comp T
  \ar[d]_<<<<<{\mu} 
&
T
  \ar[l]_-{T\eta} 
  \ar@{=}[dl]  
\\
& 
T
}
\qquad
\qquad
\xymatrix@C=3em@R=3em@M=0.5em{
T \comp T \comp T 
  \ar[r]^-{T\mu} 
  \ar[d]_-{\mu} 
& 
T \comp T
  \ar[d]^-{\mu}
\\
T \comp T 
  \ar[r]_-{\mu} 
& 
T
}
\]
The two formulations are interchangeable.
They share the same~$T$ and~$\eta$, while Kleisli extension and multiplication can be switched using
\[
  \bind f \defeq \mu_B \comp T f
  \qquad\text{and}\qquad
  \mu_A \defeq \bind {\id_{TA}}.
\]
We shall use whichever formulation is more convenient in any given case.

One part of the abstract representation of second-order functionals shall involve morphisms in the
\emph{Kleisli category} $\C_T$ of a monad $(T,\eta,\bind{(-)})$ on a category $\C$.
Recall that an object of~$\C_T$ is just an object
of~$\C$, and a morphism $f : A \to_T B$ in~$\C_T$ is a morphism $f :
A \to TB$ in~$\C$, referred to as a \emph{Kleisli map}. The Kleisli identities
$\id_A : A \to_T A$ are given by components of the unit $\eta_A : A \to TA$, and
the Kleisli composition of $f : A \to_T B$ and $g : B \to_T C$ is defined using the 
Kleisli extension, as $g \comp_T f \defeq \bind g \comp f$.

Our final category-theoretic ingredient is a comodule over a monad.
Given a monad $(T,\eta,\bind{(-)})$ on a category $\C$, we define a \emph{right
(monad) $T$-comodule} $(F,\cook)$ in a category $\D$ to be given by a
contravariant functor $F : \C^{\op} \to \D$ (the \emph{carrier})
together with a natural transformation $\cook : F \to F \comp T$ (the
\emph{structure map}), such that the following diagrams commute:
\[
  \xymatrix@C=3em@R=3em@M=0.5em{
    F 
    \ar[r]^-{\cook} 
    \ar@{=}[dr]
    & 
    F \comp T 
    \ar[d]^-{F\eta}
    \\
    & F
  }
  \qquad
  \qquad
  \xymatrix@C=3em@R=3em@M=0.5em{
    F 
    \ar[r]^-{\cook}
    \ar[d]_-{\cook}
    &
    F \comp T
    \ar[d]^-{\cook T}
    \\
    F \comp T
    \ar[r]_-{F\mu}
    &
    F \comp T \comp T
  }
\]
Note that the contravariance of $F$ causes $F \eta$ and $F \mu$ to reverse direction relative to~$\eta$ and~$\mu$. 
And to be formally precise, we should have used the opposite functor~$T^\op : \C^\op \to \C^\op$ instead of~$T$. However, since these two functors have precisely the same actions, we prefer not to sprinkle the diagrams with $\op$-superscripts.

A \emph{morphism of right $T$-comodules} $\theta : (F,\cook) \to
(F',\cook')$ is a natural transformation $\theta : F \to
F'$, such that the following diagram commutes:
\[
  \xymatrix@C=3em@R=3em@M=0.5em{
    F \comp T
    \ar[r]^-{\theta T}
    &
    F' \comp T
    \\
    F
    \ar[u]^-{\cook}
    \ar[r]_-{\theta}
    &
    F'
    \ar[u]_-{\cook'}
  }
\]
It is clear that the right $T$-comodules in~$\D$ and
morphisms between them form a category~$\CoMod{T}{\D}$. Indeed, comodule morphisms compose as the underlying
natural transformations, while the identity morphism at $(F,\cook)$ is the identity transformation $\id_F :
F \to F$.

It is admittedly a bit unusual to combine monads with comodules.
In algebra, modules go with monoids, so it stands to reason that in category theory modules go with monads~\citep{PirogWG:ModulesOverMonads}, and comodules with comonads.
Let us show that our definition is equivalent to one that tradition suggests.
We recall that a right (monad) $T$-module in a category $\E$
consists of a functor $F : \C \to \E$, together with a natural transformation
$\mathsf{m} : F \comp T \to F$, such that $\mathsf{m} \comp F\eta = \id$ and
$\mathsf{m} \comp \mathsf{m}T = \mathsf{m} \comp F\mu$.

\begin{prop}
  \label{prop:comodules-modules} 
  Given a monad $(T,\eta,\bind{(-)})$ on a category $\C$, the following are
  equivalent notions:
  \begin{enumerate}
    \item \label{prop:comodules-modules:1} $(F,\cook)$ is a right (monad)
    $T$-comodule in a category $\D$.
    \item \label{prop:comodules-modules:2} $(F,\cook)$ is a right (comonad)
    $T^\op$-comodule in a category $\D$.
    \item \label{prop:comodules-modules:3} $(F,\cook)$ is a right (monad)
    $T$-module in a category $\D^\op$.
  \end{enumerate}
\end{prop}

\begin{proof}
  The equivalence between (\ref{prop:comodules-modules:1}) and
  (\ref{prop:comodules-modules:2}) follows trivially once we put back the
  omitted $\op$-superscripts in the definition of a right monad $T$-comodule and observe that passing between monads on~$\C$ and comonads on~$\C^\op$ is
  just a matter of taking opposites, i.e.,
  a monad $(T,\eta,\mu)$ on~$\C$ corresponds to a comonad $(T^\op,\eta^\op,\mu^\op)$ on $\C^\op$.

  To establish the equivalence between (\ref{prop:comodules-modules:1}) and (\ref{prop:comodules-modules:3}),
  consider a right $T$-module in~$\D^\op$, i.e., a functor $F : \C \to \D^\op$ together
  with a natural transformation $\mathsf{m} : F \comp T \to F$ in the functor category $[\C, \D^\op]$. This is equivalently a functor $F : \C^\op \to \D$ with a natural transformation $\mathsf{m} : F \to F \comp T$, this time in the functor category
  $[\C^\op, \D]$, which are precisely the ingredients of a right $T$-comodule. The (co)module laws transform as expected, too.
\end{proof}

For morphisms, the correspondence is a bit more
involved, although in a standard way~\citep{Plotkin:TensorsOfModels}. First, based on the observations
made in the proof of \cref{prop:comodules-modules}, when relating
(\ref{prop:comodules-modules:1}) and (\ref{prop:comodules-modules:2}), we
immediately have the following result.

\begin{prop}
  The category $\CoMod{T}{\D}$ of right (monad) $T$-comodules is the same as the
  category of right (comonad) $T^\op$-comodules in $\D$, for the 
  comonad $(T^\op,\eta^\op,\mu^\op)$ on $\C^\op$
  induced by the monad $(T,\eta,\bind{(-)})$ on $\C$.
\end{prop}

Second, recalling that a morphism $\theta : (F,\mathsf{m}) \to (F',\mathsf{m}')$
of right (monad) $T$-modules is given by a natural transformation $\theta : F
\to F'$ satisfying $\theta \comp \mathsf{m} = \mathsf{m}' \comp \theta T$, and
writing $\Mod{T}{\D^\op}$ for the category of right (monad) $T$-modules in
$\D^\op$, we have the following result.

\begin{prop}
  $\CoMod{T}{\D} \cong (\Mod{T}{\D^\op})^\op$. 
\end{prop}

\begin{proof}
  This equivalence follows straightforwardly from observing that given functors
  $F, G : \C^\op \to \D$ and a natural transformation $\theta : F \to G$
  from $F$ to $G$, then this is equivalent to being given two functors $F', G' : \C
  \to \D^\op$ and a natural transformation $\theta' : G' \to F'$ from $G'$
  to $F'$.
\end{proof}

The above observations make it clear that we could also develop the rest of
the paper using either comodules of a comonad on $\Cont^\op$, or
modules of a monad in $\Type^\op$. However, we find it easier to avoid opposite
categories as much as possible, which makes right $T$-comodules of a monad the preferred choice.


\section{Comodule Representations of Second-Order Functionals}
\label{sec:comodule-representation-of-functionals}

We are now ready to recast the tree representation of a second-order functional in a category-theoretic language that will lead us to a general notion of representation in \cref{sec:comodule-representation-functional}.
Here is the relevant diagram~\eqref{dia:tree-representation} again:
\[
  \xymatrix@C=1em@R=3em@M=0.5em@L=0.5em{
    \prd{a \of A} P\, a
    \ar[rr]^-{F}
    \ar@/_1pc/[dr]_(0.25){\cook_{A,P}}
    &&
    \prd{b \of B} Q\, b
    \\
    &
    \prd{t \of \Tree{A,P}} \Path[A,P]{t}
    \ar@/_1pc/[ur]_(0.75){\lam{\alpha} \lam{b} \eat_F\, (\alpha\, (\tree_F\, b))}
  }
\]

\subsection{Tree Monad on Containers}
\label{sec:tree-monad}

The trees and paths appearing in the bottom middle dependent product in~\eqref{dia:tree-representation} arise from a monad $(\Tr,\eta,\bind{(-)})$ on $\Cont$, whose underlying functor is

\[
\Tr (\cont A P) \defeq \cont {\Tree{A,P}}
{\lam{t} \Path[A,P]{t}}.
\] 
We call it the \emph{(node-labelled) tree monad} on $\Cont$.
The definitions of the unit and Kleisli extension involve some amount of functional programming that the reader can find in \cref{appendix:definition-of-tree-monad}.
Here we satisfy ourselves with an informal description.

The component $\eta_{\cont A P} : (\cont A P) \to (\cont {\Tree{A,P}} {\lam{t} \Path{A,P}(t)})$  of the unit at $\cont A P$ is the container morphism whose shape map takes $a : A$ to the tree whose root is labelled with~$a$ and each path immediately leads to a leaf, and whose position map takes the leaf at the end of a path $\pstep(p,\lam{\_} \leaf)$ to the corresponding branch~$p : P\,a$,
pictorially illustrated as follows:
{
\setlength{\abovedisplayskip}{0pt}
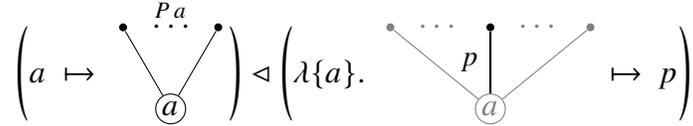
\begin{align*}
  \cont
    {\Biggl( 
      a ~ \mapsto 
      \raisebox{-16pt}{
      \begin{forest}
        for tree={%
          label/.option=content,
          grow=north,
          content=,
          circle,
          minimum size=3pt,
          inner sep=1pt
        }
        [ $a$, circle, draw
          [, fill] [$\stackrel{P\, a}{\ldots}$ , no edge] [, fill]
        ]
      \end{forest}} 
    \Biggr)}
    {\Biggl(
      \ilam{a}
      \raisebox{-16pt}{
      \begin{forest}
        for tree={%
          label/.option=content,
          grow=north,
          content=,
          circle,
          minimum size=3pt,
          inner sep=1pt,
        }
        [ \color{gray} $a$, circle, draw={gray}
          [, fill={gray}, edge={gray}] 
          [\color{gray}\ldots , no edge] 
          [, fill, edge label={node[midway,left,font=\small]{$p$}}, edge={thick}] 
          [\color{gray}\ldots , no edge] 
          [, fill={gray}, edge={gray}]
        ]
      \end{forest}} 
      \mapsto ~ p
    \Biggr)} 
\end{align*}
}
Note that neither the type $P\,a$ nor the branching of the trees need be finite.

The Kleisli extension $\bind {h} : \Tr(\cont A P) \to \Tr (\cont B Q)$ of a
container morphism $h = \cont f g : \cont A P \to \Tr(\cont B Q)$ is given by
turning an $(A,P)$-tree into a $(B,Q)$-tree, by recursively grafting a
$B$-labelled tree into every $A$-node using both~$f$ and~$g$.

In \cref{fig:tree-kleisli-extension}, we illustrate how the shape map of $\bind{h} = \cont {\shapeMap{\bind{h}}} {\positionMap{\bind{h}}}$ works on an $(A,P)$-tree $t$, whose 
root is labelled with~$a$ and the $p$-th subtree is~$t_p$.
The map $\shapeMap{\bind{h}}$ replaces the root $a$ with the $(B,Q)$-tree $f\, a$, and
grafts further trees into its leaves. Specifically,
the leaf at the end of a path $\pth{q} : \Path[B,Q]{f\, a}$ is replaced by the recursively constructed
subtree $\shapeMap{\bind{h}}\, (t_{g\, \pth{q}})$. Recursion makes sure that the process is repeated throughout~$t$, until the leaves of $t$ are reached and left unchanged.
%

\begin{figure}[h]
    \centering
\small
\begin{align*}
  \shapeMap{\bind{h}}\, t 
  \quad \defeq \quad &
  \shapeMap{\bind{h}} \Biggl( 
      \raisebox{-25pt}{
      \begin{forest}
        for tree={%
          label/.option=content,
          grow=north,
          content=,
          minimum size=3pt,
          inner sep=1pt,
        }
        [ $a$, circle, draw
          [\ldots, tria ] 
          [\ldots , no edge, before computing xy={s+=0.5em}] 
          [\;$t_{p}$\; , tria, edge label={node[midway,left,font=\small]{$p$}}, edge={thick}] 
          [\ldots , no edge, before computing xy={s+=0.25em}] 
          [\ldots, tria ]
        ]
      \end{forest}} 
  \Biggr)
  \\[10pt]
  \quad \defeq \quad &
  \raisebox{-25pt}{
    \begin{forest}
      for tree={%
        label/.option=content,
        grow=north,
        content=,
        minimum size=3pt,
        inner sep=1pt,
      }
      [ $f\, a$, tria
        [$\shapeMap{\bind{h}}\, \ldots$, tria ] 
        [\ldots , no edge, before computing xy={s+=3em}] 
        [\;$\shapeMap{\bind{h}}\, (t_{g\, \pth{q}})$\; , tria, edge label={node[midway,left,font=\small]{$\pth{q}$}}, edge={thick}] 
        [\ldots , no edge, before computing xy={s+=1.5em}] 
        [$\shapeMap{\bind{h}}\, \ldots$, tria ]
      ]
    \end{forest}}
\end{align*}
  \caption{The Kleisli extension $\bind{h}$ of a container morphism $h = \cont f
  g : \cont A P \to \Tr(\cont B Q)$ at a tree $t : \Tree{A,P}$, whose root is labelled
  with~$a$ and the $p$-th subtree of the root node is~$t_p$.}
  \label{fig:tree-kleisli-extension}
\end{figure}

For every $(A,P)$-tree $t$, the position map $\positionMap{\bind{h}}$
maps a composite path in the $(B,Q)$-tree $\shapeMap{\bind{h}}\, t$ to a 
path in the $(A,P)$-tree~$t$. The process starts at the
roots of~$t$ and $\shapeMap{\bind{h}}\, t$, and proceeds by stepping through the two
trees while mapping every composite path $\pth{q} : \Path[B,Q]{f\, a}$ in
$\shapeMap{\bind{h}}\, t$'s subtrees of the form $f\, a$, for some $a : A$, to single
path links $g\, \pth{q}$ (at the given node~$a$) in the original $(A,P)$-tree~$t$.

\begin{prop}
  \label{prop:tree-monad}
  $(\Tr,\eta,\bind{(-)})$, as defined in \cref{appendix:definition-of-tree-monad}, forms a monad on $\Cont$.
\end{prop}

For proving \cref{prop:tree-monad}, it is useful to note that $\Tr (\cont A P)$ can
be more abstractly constructed as the initial algebra of an endofunctor on
$\Cont$ given by
\[
  \cont B Q 
  \ \mapsto \ 
  \cont 
    {\big(\One + \sm{a \of A} (P\, a \to B)\big)} 
    {\big[\star \mapsto \One , (a,v) \mapsto \sm{p \of P\, a} Q\, (v\, p)\big]},
\] 
which is written more succinctly using the categorical structure of~$\Cont$ as
\[
  \cont B Q
  \ \mapsto \ 
  \cSum {\cId} {\cComp {(\cont A P)} {(\cont B Q)}}.
\]
A detailed proof that this yields a monad on can be found in
\cite[Theorem~4.5]{GambinoK:PolyMonads}.
We reckon the abstract viewpoint could be useful for future generalisations of
comodule representations, e.g., for combining different monads and comodules.

\subsection{(Kleisli) Container Morphism}

The objects of the form $\prd{a \of A} P\, a$ in the tree representation of second-order functionals~\eqref{dia:tree-representation} are isomorphic to the cointerpretation of containers~$\cont A P$ at~$\One$ because $\cosem{\cont A P}\, \One = \prd{a \of A} (P\, a \times \One) \cong \prd{a \of A} P\, a$,
a viewpoint that provides a useful perspective.
We shall only ever care about the cointerpretation at~$\One$, so hereon we repurpose the notation $\cosem{-}$ for the functor
$\Cont^\op \to \Type$ defined by
\begin{equation*}
  \cosem{\cont A P} \defeq \prd{a \of A} P\, a
  \quad\text{and}\quad
  \cosem{\cont f g} \defeq \lam{\alpha} \lam{b} g\, (\alpha\, (f \, b)).
\end{equation*}
With this notation, the map $\lam{\alpha} \lam{b} \eat_F\, (\alpha\, (\tree_F\, b)) : \prd{t \of \Tree{A,P}} \Path[A,P]{t} \to \prd{b \of B} Q\, b$ appearing in a tree representation~\eqref{dia:tree-representation} may be written more succinctly as
\[
  \cosem{\cont {\tree_F} {\eat_F}} 
    : \cosem{\Tr(\cont{A}{P})} \to \cosem{\cont B Q},
\]
where
$
  \cont {\tree_F} {\eat_F} : \cont B Q \to \Tr(\cont A P)
$
is a morphism in the Kleisli category of the monad $\Tr$.
This allows us to recast diagram~\eqref{dia:tree-representation} more abstractly as follows:
\begin{equation}
  \label{dia:tree-representation-more-abstract}
  \begin{gathered}
  \xymatrix@C=1em@R=3em@M=0.5em@L=0.5em{
    \cosem{\cont A P}
    \ar[rr]^-{F}
    \ar@/_1pc/[dr]_(0.25){\cook_{\cont A P}}
    &&
    \cosem{\cont B Q}
    \\
    &
    \cosem{\Tr(\cont{A}{P})}
    \ar@/_1pc/[ur]_(0.75){\cosem{\cont {\tree_F} {\eat_F}}}
  }
\end{gathered}
\end{equation}
For everything to fall into place, notice that~$\cook_{\cont A P}$ has the right shape and definition to be a component of the structure map of a right $\Tr$-comodule.

\begin{prop}
  \label{prop:tree-monad-comodule-map}
  The map $\cook_{\cont A P} : \cosem{\cont A P} \to \cosem{\Tr(\cont{A}{P})}$ 
  appearing in a tree representation~\eqref{dia:tree-representation-more-abstract} is
  a component at $\cont A P$ of a structure map $\cook : \cosem{-} \to \cosem{\Tr(-)}$ of
  a right $\Tr$-comodule with carrier $\cosem{-} : \Cont^\op \to \Type$.
\end{prop}

\subsection{Comodule Representations of Second-Order Functionals}
\label{sec:comodule-representation-functional}

The category-theoretic formulation~\eqref{dia:tree-representation-more-abstract} of tree representation of second-order functionals immediately suggests the following general definition.

\begin{defn}
  \label{def:comodule-representation}
  Let $(T,\eta,\bind{(-)})$ be a monad on~$\Cont$ and ${\cook : \cosem{-} \to
  \cosem{T(-)}}$ a right $T$-comodule structure map on $\cosem{{-}}$.
  A \emph{$(T,\cosem{-},\cook)$-representation} of a second-order functional ${F :
  \left(\prd{a \of A} P\, a\right) \to \left(\prd{b \of B} Q\, b\right)}$ is
  a container morphism $\cont {\tree_F} {\eat_F} : \cont B Q \to
  T(\cont A P)$ such that the following diagram commutes:
  \begin{equation}
    \label{dia:comodule-representation}
    \begin{gathered}
    \xymatrix@C=1em@R=3em@M=0.5em@L=0.5em{
      \cosem{\cont A P}
      \ar[rr]^-{F}
      \ar@/_1pc/[dr]_(0.25){\cook_{\cont A P}}
      &&
      \cosem{\cont B Q}
      \\
      &
      \cosem{T(\cont{A}{P})}
      \ar@/_1pc/[ur]_(0.75){\cosem{\cont {\tree_F} {\eat_F}}}
    }
    \end{gathered}
  \end{equation}
  A second-order functional for which there exists\footnote{A type-theoretic clarification: we mean existence in the sense of the logical existential quantifier~$\exists$, as opposed to dependent sum~$\Sigma$. That is, representability is a property, not a structure.} such a representation is said to be \emph{$(T, \cosem{-}, \cook)$-representable}, or just \emph{(comodule) representable}.
\end{defn}

For the rest of the section we shall be concerned with comodule representation for a fixed monad $(T,\eta,\bind{(-)})$ and a right $T$-comodule $\cook : \cosem{-} \to \cosem{T(-)}$.

\begin{thm}
  \label{thm:comodule-representable-functionals-category}
  The comodule-representable second-order functionals form a subcategory
  $\RFun{T,\cosem{-},\cook}$ of the category~$\Fun$ of all second-order
  functionals.
\end{thm}

\begin{proof}
  To be precise, the objects of $\Fun$ are containers $\cont A P$ and morphisms
  are maps $\cosem{\cont A P} \to \cosem{\cont B Q}$. The objects of
  $\RFun{T,\cosem{-},\cook}$ are containers $\cont A P$ and the morphisms $F :
  \cont A P \to \cont B Q$ are the comodule representable functionals $F :
  \cosem{\cont A P} \to \cosem{\cont B Q}$. We must show that the identity maps
  are representable and that the representable functionals are closed under
  composition.

  For any $\cont A P$, the \emph{identity functional} $\id_{\cosem{\cont A P}} :
  \cosem{\cont A P} \to \cosem{\cont A P}$ is representable by the
  component $\eta_{\cont A P} : \cont A P \to T(\cont A P)$ of the
  unit for~$T$.

  If $F : \cosem{\cont A P} \to \cosem{\cont B Q}$ and $G : \cosem{\cont B Q} \to \cosem{\cont C R}$
  are represented by $\cont{\tree_F} {\eat_F}$ and $\cont {\tree_G} {\eat_G}$, respectively,
  then their composition $G \comp F$ is represented by the Kleisli composition of their
  representations, because the next diagram commutes:
  
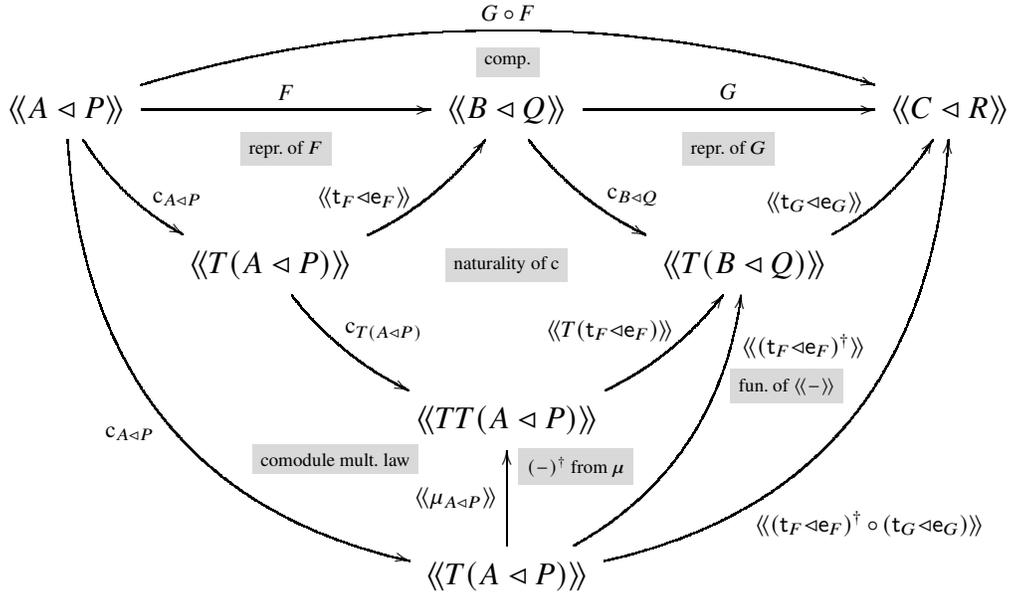
\begin{equation*}
    \xymatrix@C=1em@R=3em@M=0.5em@L=0.3em{
      \cosem{\cont A P}
      \ar@/^2.5pc/[rrrr]^-{G \,\comp\, F}
      \ar@{}@<27pt>[rrrr]_-{\dcomment{comp.}}
      \ar[rr]^-{F}
      \ar@/_1pc/[dr]^>>>>{\cook_{\cont A P}}
      \ar@/_4.5pc/[dddrr]_-{\cook_{\cont A P}}
      \ar@{}@<-5pt>[rr]_-{\dcomment{repr.~of $F$}}
      &&
      \cosem{\cont B Q}
      \ar[rr]^-{G}
      \ar@/_1pc/[dr]^>>>>{\cook_{\cont B Q}}
      \ar@{}@<-5pt>[rr]_-{\dcomment{repr.~of $G$}}
      &&
      \cosem{\cont C R}
      \\
      &
      \cosem{T(\cont{A}{P})}
      \ar@/_1pc/[ur]^<<<<{\cosem{\cont {\tree_F} {\eat_F}}}
      \ar@/_1pc/[dr]^-{\cook_{T(\cont A P)}}
      \ar@{}[rr]|-{\dcomment{naturality of $\cook$}}
      &&
      \cosem{T(\cont B Q)}
      \ar@/_1pc/[ur]^<<<{\cosem{\cont {\tree_G} {\eat_G}}}
      \\
      &&
      \cosem{TT(\cont A P)}
      \ar@/_1pc/[ur]^-{\cosem{T(\cont {\tree_F} {\eat_F})}}
      \\
      &&
      \cosem{T(\cont A P)}
      \ar@/_4.5pc/[uuurr]_(0.35){\quad\;\;\cosem{\bind{(\cont {\tree_F} {\eat_F})} \,\comp\, (\cont {\tree_G} {\eat_G})}}
      \ar[u]^-{\cosem{\mu_{\cont A P}}}
      \ar@/_2pc/[uur]_>>>>>>{\cosem{\bind{(\cont {\tree_F} {\eat_F})}}}
      \ar@{}@/_2pc/[uur]_<<<<<<<<<<<<<<<<<<<<<<<{\dcomment{fun.~of $\cosem{-}$}}
      \ar@{}@<30pt>[u]^(0.75){\dcomment{comodule mult.~law}}
      \ar@{}@<-1pt>[u]_>>>{\dcomment{$\bind{(-)}$ from $\mu$}}
    }
    \qedhere
  \end{equation*}
\end{proof}

\noindent
We proved more than was asked, namely that there is a functor
\begin{equation*}
  \F : \Cont_T^\op \to \RFun{T,\cosem{-},\cook}, 
\end{equation*}
which acts trivially on objects and takes a Kleisli container morphism $\cont f g : \cont B Q \to T(\cont A P)$ to the second-order functional represented by it,
\[
  \F(\cont f g) \defeq \cosem{\cont f g} \comp \cook_{\cont A P} :
  \cosem{\cont A P} \to \cosem{\cont B Q}.
\]
The functor is obviously full, and it also preserves finite products:

\begin{prop}
  The categories $\Cont_T^\op$ and $\RFun{T,\cosem{-},\cook}$ have finite products, which $\F$ preserves.
\end{prop}

\begin{proof}
  It is a general fact that a Kleisli category for a monad has finite coproducts when the underlying category does~\citep{Szigeti:LimitsColimitsInKleisliCat}.
  Thus $\Cont_T^\op$ has finite products because in \cref{sec:containers-morphisms} we
  showed that~$\Cont$ has finite coproducts.
  Explicitly, the products in~$\Cont_T^\op$ are the coproducts in~$\Cont$, except that the coproduct injections of~$\Cont$ must be post-composed with the unit of the monad.

  The finite products in $\RFun{T,\cosem{-},\cook}$ also arise from the fact that $\Cont$ has finite coproducts, and that $\cosem{-}$ maps them to products, as we showed in~\cref{sec:interp-cointerp}. Let us spell them out to verify that all the structure is represented.

  The terminal object in $\RFun{T,\cosem{-},\cook}$ is the initial object $\cont \Zero {\lam{\_}
  \Zero}$ in $\Cont$. For any $\cont A P$, the unique functional
  $!^\text{r}_{\cont A P}$ into the terminal object is defined as in the following diagram,
  which also shows that it is represented:
  \begin{align*}
    \xymatrix@C=1em@R=3em@M=0.5em@L=0.5em{
      \cosem{\cont A P}
      \ar[rr]^-{!^\text{r}_{\cont A P}}
      \ar@/_2.5pc/[ddr]_(0.35){\cook_{\cont A P}}
      \ar@/_1pc/[dr]^-{!_{\cosem{\cont A P}}}
      \ar@{}@<-15pt>[rr]_-{\dcomment{def.}}
      &&
      \cosem{\cont \Zero {\lam{\_} \Zero}}
      \\
      & 
      \One
      \ar@/_1pc/[ur]^-{\cong}
      \\
      &
      \cosem{T(\cont{A}{P})}
      \ar@/_2.5pc/[uur]_(0.65){\cosem{?^\tinyc_{T(\cont A P)}}}
      \ar[u]_-{!_{\cosem{T(\cont A P)}}}
      \ar@{}@<1pt>[u]^(0.8){\dcomment{univ.~pr.~of $\One$}}
      \ar@{}@<-5pt>[u]_(0.8){\dcomment{$\cosem{-}$ pres.~$\One$}}
    }
  \end{align*}
  The uniqueness of $!^\text{r}_{\cont A P}$ follows from that of $!_{\cosem{\cont A P}}$.

  The product $\rProd{(\cont A P)}{(\cont B Q)}$ of $\cont A P$ and $\cont B Q$ in $\RFun{T,\cosem{-},\cook}$
  is given by the coproduct  $\cSum{(\cont A P)}{(\cont B Q)}$ in $\Cont$.
  The first projection $\rFst$ is defined by the following diagram, which also shows
  that it is representable:
  
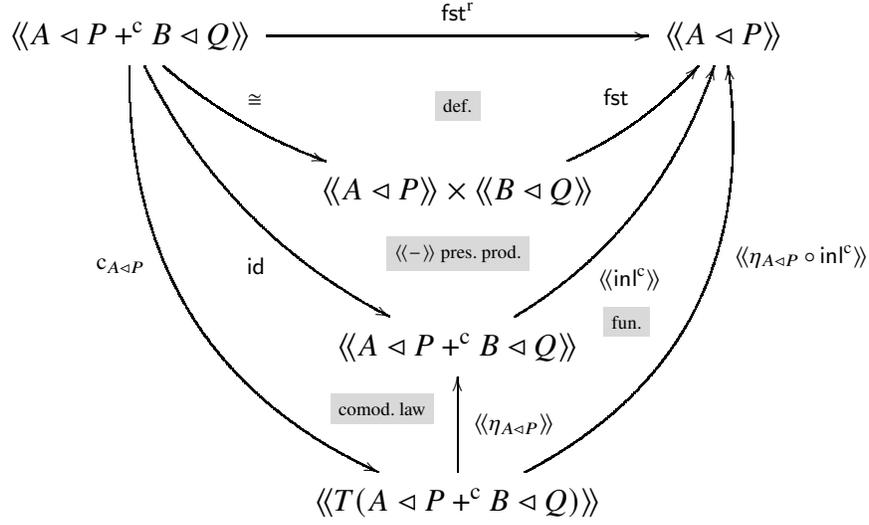
\begin{align*}
    \xymatrix@C=1em@R=3em@M=0.5em@L=0.5em{
      \cosem{\cSum{\cont A P}{\cont B Q}}
      \ar[rr]^-{\rFst}
      \ar@/_1pc/[dr]^-{\cong}
      \ar@/_3.5pc/[dddr]_(0.35){\cook_{\cont A P}}
      \ar@/_1.5pc/[ddr]_(0.55){\id}
      \ar@{}@<-15pt>[rr]_-{\dcomment{def.}}
      \ar@{}@<-70pt>[rr]_-{\dcomment{$\cosem{-}$ pres.~prod.}}
      &&
      \cosem{\cont A P}
      \\
      &
      \cosem{\cont A P} \times \cosem{\cont B Q}
      \ar@/_1pc/[ur]^-{\fst}
      \\
      &
      \cosem{\cSum{\cont A P}{\cont B Q}}
      \ar@/_1.5pc/[uur]_(0.4){\!\!\!\!\cosem{\cInl}}
      \ar@{}@<-25pt>[uur]_(0.30){\dcomment{fun.}}
      \\
      &
      \cosem{T(\cSum{\cont A P}{\cont B Q})}
      \ar@/_3.5pc/[uuur]_(0.65){\cosem{\eta_{\cont A P} \,\comp\, \cInl}}
      \ar[u]_-{\cosem{\eta_{\cont A P}}}
      \ar@{}@<3pt>[u]^(0.6){\dcomment{comod.~law}}
    }
  \end{align*}
  The second projection $\rSnd$ is defined analogously.

  The pairing $\rPair F G$ of second-order functionals $F : \cosem{\cont A P} \to
  \cosem{\cont B Q}$ and $G : \cosem{\cont A P} \to \cosem{\cont C R}$,
  represented respectively by $\cont {\tree_F} {\eat_F}$ and
  $\cont {\tree_G} {\eat_G}$, is given by the following diagram,
  which also shows that the pairing is representable:
  
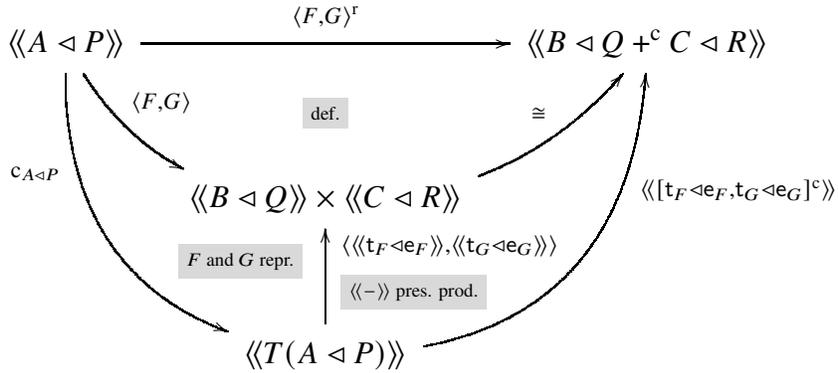
\begin{align*}
    \xymatrix@C=1em@R=3em@M=0.5em@L=0.5em{
      \cosem{\cont A P}
      \ar[rr]^-{\rPair F G}
      \ar@/_3pc/[ddr]_(0.25){\cook_{\cont A P}}
      \ar@/_1.5pc/[dr]^-{\!\!\!\pair F G}
      \ar@{}@<-15pt>[rr]_-{\dcomment{def.}}
      &&
      \cosem{\cSum{\cont B Q}{\cont C R}}
      \\
      &
      \cosem{\cont B Q} \times \cosem{\cont C R}
      \ar@/_1.5pc/[ur]^-{\cong}
      \\
      &
      \cosem{T(\cont{A}{P})}
      \ar@/_3.5pc/[uur]_(0.75){\cosem{\cCopair{\cont {\tree_F} {\eat_F}}{\cont {\tree_G} {\eat_G}}}}
      \ar[u]_(0.7){\pair{\cosem{\cont {\tree_F} {\eat_F}}}{\cosem{\cont {\tree_G} {\eat_G}}}}
      \ar@{}@<3pt>[u]^(0.6){\dcomment{$F$ and $G$ repr.}}
      \ar@{}[u]_(0.4){\dcomment{$\cosem{-}$ pres. prod.}}
    }
  \end{align*}
  Note that the definition of the pairing does not depend on the representations.
  The required universal property laws of projections follows from the universal properties of products in $\Type$ and
  coproducts in $\Cont$.

  That $\F$ preserves finite products is immediate from its definition
  and the fact that $\cosem{-}$ maps coproducts to products.
\end{proof}

\subsection{Algebra Representations of Second-Order Functionals}
\label{sec:algebra-representations-of-functionals}

In this section we discuss an alternative, but equivalent, way of representing second-order functionals, namely by extension maps for monad algebras on the identity container, as given in \cref{def:algebra-representation} below.

Firstly, note that the product-forming functor $\cosem{-} : \Cont^\op \to \Type$
is represented by the identity container~$\cId$, i.e., $\cosem{-}$ is naturally isomorphic to $\Cont(- , \cId)$. Indeed, a morphism $\cont A P \to \cId$ consists of a trivial map $A \to \One$, which may be discarded, and an element of  $\prd{a \of A} (\One \to P\, a)$, which is isomorphic to $\prd{a \of A} P\,a$.
For the rest of the section, we identify $\cosem{\cont A P}$ and $\Cont(\cont A P , \cId)$.

Consider a monad $(T,\eta,\bind{(-)})$ on $\Cont$.
Based on the above, the
structure map $\cook : \cosem{-} \to \cosem{T(-)}$ of a $T$-comodule on~$\cosem{-}$
corresponds to a natural transformation $\cook_{} : \Cont(- , \cId) \to \Cont(T(-) , \cId)$,
which by the Yoneda lemma is represented by a unique container morphism $\alg :
T(\cId) \to \cId$.
Furthermore, the comodule laws for~$c$ translate through the Yoneda lemma
to laws stating that $\alg$ constitutes a monad $T$-algebra on~$\cId$.

We may also pass in the opposite direction. A monad $T$-algebra ${\alg : T(\cId) \to \cId}$
induces at each $\cont A P$ a map $\cook_{\cont A P} : \cosem{\cont A P} \to \cosem{T (\cont A P)}$, given by the algebra extension operation defined as $\cook_{\cont A P}\, h \defeq \alpha \circ T\,h$. The monad algebra laws guarantee that $\cook_{\cont A P}$ is the component of a comodule structure map.

We need one more general observation:
a monad $T$-algebra $\alpha : T X \to X$ for a monad $(T,\eta,\bind{(-)})$ on a category $\C$
induces an operation taking a Kleisli map $f : A \to T B$ to the map $\alpha \circ T ({-}) \circ f : \C(B,X) \to \C(A,X)$.

\begin{defn}
  \label{def:algebra-representation}%
  Let $\alpha : T(\cId) \to \cId$ be an algebra for a monad $(T,\eta,\bind{(-)})$ on $\Cont$.
  A second-order functional
  $F : \left(\prd{a \of A} P\, a\right) \to \left(\prd{b \of B} Q\, b\right)$,
  construed as 
  \[
  F : \Cont(\cont A P , \cId) \to \Cont(\cont B Q , \cId),
  \]
is \emph{$(T,\alpha)$-representable}, or just \emph{algebra representable}, if
there is a container map $\cont {\tree_F} {\eat_F} : \cont B Q \to
T(\cont A P)$, such that $F = \alpha \circ T ({-}) \circ (\cont {\tree_F} {\eat_F})$.
\end{defn}

The reader may verify that the correspondence between $T$-comodules and monad $T$-algebras on~$\cId$, as described above, does indeed mediate between our original \cref{def:comodule-representation} and the above one.

\Cref{def:algebra-representation} provides a useful perspective.
For instance, it took a little thinking to discover the comodule structure map for the tree monad~$\Tr$
from \cref{prop:tree-monad-comodule-map}, whereas the same example expressed in terms of algebra representations is automatic: as we observed in the discussion following \cref{prop:tree-monad}, $\Tr$ is the free monoid monad for the monoidal structure induced by $\cComp{}{}$, so one only needs to notice that $\cId$ is clearly a monoid for~$\cComp{}{}$, hence it carries a $\Tr$-algebra structure.

To our minds, the comodule representations are more directly motivated by the computational perspective on the topic, whereas the algebra representations are quite natural from a category-theoretic point of view.
In the rest of the paper, we shall take comodule representations as primary, and refer to the algebra representations when they provide an additional insight.


\section{Examples of Comodule Representations}
\label{sec:examples}

In this and the next sections, we demonstrate that comodule representations also
arise in many other situations, apart from tree representations.
Once again, we begin with a concrete example that suggests a general construction in \cref{sec:monads-from-mendler-style-algebras}.

\subsection{Finite Support}
\label{sec:finite-support}

If $F : \left(\prd{a \of A} P\, a\right) \to \left(\prd{b \of B} Q\, b\right)$ has a representation $\cont {\tree_F} {\eat_F}$, then for all arguments~$h : \prd{a \of A} P\, a$ and~$b : B$, the value $F \, h \, b$ may be computed by querying~$h$ at only finitely many values of~$h$, namely the labels appearing on the path $\cook_{A,P} \, h \, (\tree_F \, b)$.
Which labels these are depends both on~$h$ and~$b$, and they are linearly ordered by their placement on the path.
An alternative notion of representation arises when we still require that values $F \, h \, b$ depend on finitely many labels, but they may depend only on~$b$, and are not ordered.

Let us work with types as sets.
Say that $F$ has \emph{finite support} when for every $b : B$ there is a finite
subset $A_{F,b} \subseteq A$ such that $F \, h \,b$ can be computed already from the restriction $h{\restriction}_{A_{F,b}} : \prd{a \of A_{F,b}} P a$, i.e., if $h$ and $h'$ agree on~$A_{F,b}$ then $F \, h \, b = F \, h' \, b$.
The identity functional $\id : \left(\prd{a \of A} P\, a\right) \to \left(\prd{a \of A} P\, a\right)$ is finitely supported by singletons $A_{\id,a} = \finset{a}$, and because finite sets are closed under finite unions, finitely supported functionals are closed under composition.

This turns out to be a case of a comodule representation. The underlying monad $(T,\eta,\bind{(-)})$  on containers is defined as
\begin{equation*}
  T(\cont A P) \defeq \cont {\FPow{A}} {\malg{P}},
\end{equation*}
where $\FPow$ is the finite powerset functor (itself a monad on sets, whose unit forms singletons and whose multiplication is given by the union of sets), and the operation $\malg{P} : \FPow{A} \to \Type$ 
forms dependent products $\malg{P}\, S \defeq \prd{a \of S} P\, a$.
The component of the unit $\eta_{\cont A P} : \cont A P \to T(\cont A P)$ is the container morphism given by
\begin{equation*}
  \eta_{\cont A P} \defeq 
    \cont
    {\big(\lam{(a \of A)} \finset{a}\big)}
    {\big(\ilam{a \of A} \lam{(u \of \prd{a' \of \finset{a}} P\, a')} u\, a\big)},
\end{equation*}
where the shape map is simply the unit of the finite powerset monad $\FPow{}$ at $A$.
For any container morphism $\cont f g : \cont A P \to T(\cont B Q)$, the Kleisli extension $\bind
{(\cont f g)} : T(\cont A P) \to T(\cont B Q)$ is the container morphism given by
\begin{equation*}
  \cont
    {f^\ddagger}
    {\left(
       \ilam{S \of \FPow{A}}
       \lam{(u \of \prd{b' \of f^\ddagger S} Q\, b')}
       \lam{(a' \of S)}
          g\, \ia{a'}\, (u{\restriction}_{f\, a'})
     \right)}, 
\end{equation*}
where $f^\ddagger : \FPow{A} \to \FPow{B}$ is the Kleisli extension of $f : A \to \FPow{B}$ with respect to the finite powerset monad $\FPow$, given by $f^\ddagger S \defeq \bigcup_{a \of S} f\,a$.

The monad laws hold for the shape maps because there they coincide with the finite powerset monad.
It is not hard to check them on positions either.
The first unit law $\bind{\eta_{\cont A P}} = \id_{T(\cont A P)}$ amounts to showing,
for all $S : \FPow A$, that
\begin{equation*}
  \lam{(u : \prd{a'' \of \finset{-}^\ddagger S} P\, a'')} u{\restriction}_{S} 
  = 
  \lam{(u : \prd{a'' \of S} P\, a'')} u ,
\end{equation*}
which holds by extensionality and the monad law $\finset{-}^\ddagger S = S$.
The second unit law $\bind{(\cont f g)} \comp \eta_{\cont A P} = \cont f g$,
amounts to showing, for all $a : A$, that
\[
  \lam{(u \of \prd{b' \of f^\ddagger \finset a} Q\, b')} g\, \ia a \, (u{\restriction}_{f\, a}) 
  = 
  \lam{(u \of \prd{b' \of f a} Q\, b')} g\, \ia a \, u , 
\]
which holds by extensionality and the monad law $f^\ddagger \finset a = f\, a$. 
The associativity law follows much the same way, so we omit the details.

We also need a right $T$-comodule structure map. At a container $\cont A P$, its component $\cook_{\cont A P} : \cosem{\cont A P} \to \cosem{T(\cont A P)}$ is just the restriction operation:
\begin{align*}
  \cook_{\cont A P} &\defeq
    \lam{\left(h \of \prd{a \of A} P\, a\right)}
    \lam{\left(S \of \FPow{A}\right)}
       h{\restriction}_{S}.
\end{align*}
The naturality of~$\cook$ is established by unfolding the definitions, so we omit it here.
Next, the comodule unit law $\cosem{\eta_{\cont A P}} \comp \cook_{\cont A P} =
\id_{\cosem{\cont A P}}$ amounts to showing
\[
  \lam{(a \of A)} h{\restriction}_{\finset{a}}\, a = h,
\]
which holds by extensionality and $h{\restriction}_{\finset{a}}\, a = h\, a$.
Finally, the multiplication law $\cook_{T(\cont A P)} \comp \cook_{\cont A P} =
\cosem{\mu_{\cont A P}} \comp \cook_{\cont A P}$ amounts to showing, for all $\mathcal{S} : \FPow{(\FPow A)}$, that
\[
  (\lam{(S' \of \FPow A)} h{\restriction}_{S'}){\restriction}_{\mathcal{S}} 
  =
  h{\restriction}_{\bigcup \mathcal{S}}, 
\]
which holds by extensionality and how unions interact with restrictions.

Let us check that the given comodule does indeed yield representations of finitely supported functionals.
If $F : \left(\prd{a \of A} P\, a \right) \to
\left(\prd{b \of B} Q\, b \right)$ is represented by a container morphism $\cont
{\tree_F} {\eat_F} : \cont B Q \to \cont {\FPow A} {\left(\lam{S} \prd{a \of S} P\, a\right)}$, 
then for each $h : \prd{a \of A} P \, a$ and $b : B$, we have the following sequence of equations:
\begin{equation*}
  F \, h \, b =
  \eat_F (\cook_{A,P} \, h \, (\tree_F \, b)) =
  \eat_F (h{\restriction}_{\tree_F \, b}).
\end{equation*}
This is just what we expected: $\eat_F$ computes $F\, h \, b$ from the restriction of~$h$ to the finite subset
$\tree_F \, b : \FPow{A}$, which indeed depends only on~$b$. Also, no particular order of queries to~$h$ is imposed.

The relationship between tree-represented and finitely-supported functionals is significant both from a topological and computability point of view.
The topological observation is that a functional $\Nat^\Nat \to \Nat^\Nat$ is point-wise continuous when it has a tree representation, and uniformly continuous when it is finitely-supported.
And also, the relationship between the two notions is akin to that between Turing reductions and truth-table reductions in computability theory~\cite{Hartley:TheoryOfRecursiveFunctions}.
Indeed, in a Turing reduction the next query to the oracle may depend on the previous oracle answers, whereas a truth-table reduction must present a finite set of queries upfront.
We leave a study of comodule representability in computability theory for another day.

If so desired, we could have avoided working with sets by replacing them with a type-theoretic setup that is equipped with a suitable notion of finite subtypes~\cite{UustaluV:Finiteness,spiwack10:_const_finit} and the analogue of the finite powerset monad.
Such structures can be constructed in several ways, for instance using quotient types~\cite{Li:QuotientTypes}.

\subsection{Monads Induced by Weak Mendler-Style Algebras on a Universe}
\label{sec:monads-from-mendler-style-algebras}

The monad representing finitely supported functionals has a particular form, namely,
it acts on the shapes~$A$ of a container $\cont A P$ with a monad $\FPow{}$ on~$\Type$, and extends the positions $P : A \to \Type$ to a family $\malg P : \FPow{A} \to \Type$ in a way that is compatible with the structure of $\FPow{}$.
This turns out to be a general phenomenon that is the source of further examples of comodule representations.

The crucial observation is that the passage from the family $P : A \to \Type$ to $\malg P : \FPow{A} \to \Type$ arises through a certain algebra structure for~$\FPow{}$ on $\Type$. To motivate it, we first recall a simpler standard version of these kinds of algebras.

\begin{defn}
  Given a monad $(M,\eta,\bind{(-)})$ on a category $\C$,
  a \emph{Mendler-style $M$-algebra} $(A, \malg{(-)})$, also called an \emph{$M$-algebra in
    no-iteration form}~\cite{Marmolejo:NoIterationMonads,UustaluV:MendlerStyleAlgebras}, is given
  by an object $A$ of $\C$ (the carrier) and an (algebra) extension operation
  taking any map $f : B \to A$ to a map $\malg{f} : M B \to A$, such that
  \begin{equation*}
    \malg f \comp \eta_B = f
    \qquad\text{and}\qquad
    \malg f \comp \bind g = \malg {(\malg f \comp g)},
  \end{equation*}
  for all $f : B \to A$ and $g : C \to M B$.
\end{defn}

\noindent

Now, the extension $\malg P : \FPow{A} \to \Type$ of $P : A \to \Type$ in the previous section looks
like a case of Mendler-style $\FPow{}$-algebra on~$\Type$. However, $\Type$ has
additional structure, it is a category, which should be taken into account. In particular, the
extension operation ought to act functorially on maps between type families.
The following definition accomplishes this, and simultaneously weakens the algebraic
laws to lax ones, and allows the algebra carrier to be any type universe.

\begin{defn}
  \label{def:weak-mendler-style-T-algebra}%
Given a monad $(M,\eta,\bind{(-)})$ on $\Type$,
a \emph{weak Mendler-style $M$-algebra} on a universe~$\U$ is given by:
\begin{itemize}
\item an \emph{extension operation} taking a family $P : A \to \U$ to $\malg{P} : M A \to \U$, 
\item an \emph{action} taking a family map
  $h : \iprd{a \of A} (P\, a \to Q\, a$)
  to a family map
  $\malgfun{h} : \iprd{m \of M A} (\malg{P}\, m \to \malg{Q}\, m)$,
  where $P, Q : A \to \U$, and
\item families of maps laxly witnessing the Mendler-style $M$–algebra laws:
  \begin{itemize}
  \item $\i_P : \iprd{a \of A} (\malg P\,(\eta_A\, a) \to P\, a)$, 
    for all $P : A \to \U$, and
  \item $\j_{Q,f} : \iprd{m \of M A} (\malg Q\, (\bind f\, m) 
    \to \malg {(\malg Q \comp f)}\, m)$, 
    for all $Q : B \to \U$ and ${f : A \to M B}$,
  \end{itemize}
\end{itemize}
satisfying the following diagrammatic conditions:
\begin{itemize}
\item the action is functorial:
\[
  \xymatrix@C=5em@R=3em@M=0.5em@L=0.5em{
    \malg{P}\, m
    \ar@{=}[d]
    \ar[r]^-{\malgfun{\ilam{a} \id_{P a}}}
    &
    \malg{P}\, m
    \ar@{=}[d]
    \\
    \malg{P}\, m
    \ar[r]_-{\id_{\malg{P} m}}
    &
    \malg{P}\, m
  }
  \qquad
  \xymatrix@C=3em@R=3em@M=0.5em@L=0.5em{
    \malg{P}\, m
    \ar@{=}[d]
    \ar[rr]^-{\malgfun{\ilam{a} h' \, \comp \, h\, \ia{a}}}
    &
    &
    \malg{R}\, m
    \ar@{=}[d]
    \\
    \malg{P}\, m
    \ar[r]_-{\malgfun{h}}
    &
    \malg{Q}\, m
    \ar[r]_-{\malgfun{h'}}
    &
    \malg{R}\, m
  }
\]
\item the maps $\i_P$ and $\j_{Q,f}$ are natural with respect to $\malgfun{-}$:
\[
  \xymatrix@C=3em@R=3em@M=0.5em@L=0.5em{
    \malg{P}\, (\eta_A\, a)
    \ar[d]_-{\malgfun{h}}
    \ar[r]^-{\i_P}
    &
    P\, a
    \ar[d]^-{h}
    \\
    \malg{P'}\, (\eta_A\, a)
    \ar[r]_-{\i_{P'}}
    &
    P'\, a
  }
  \quad
  \xymatrix@C=3em@R=3em@M=0.5em@L=0.5em{
    \malg{Q}\, (\bind{f}\, m)
    \ar[d]_-{\malgfun{k}}
    \ar[r]^-{\j_{Q,f}}
    &
    \malg{(\malg{Q} \comp f)}\, m
    \ar[d]^-{\malgfun{\ilam{a} \malgfun{k}\, \ia{f\, a}}}
    \\
    \malg{Q'}\, (\bind{f}\, m)
    \ar[r]_-{\j_{Q',f}}
    &
    \malg{(\malg{Q'} \comp f)}\, m
  }
\]
\item the maps $\i_P$ and $\j_{Q,f}$ are coherent with respect to the monad laws:
\begin{gather*}
  \xymatrix@C=3em@R=3em@M=0.5em@L=0.5em{
    \malg{Q}\, (\bind{f}\, (\eta_A\, a))
    \ar@{=}[dr]_-{\text{mon.~law}~~}
    \ar[r]^-{\j_{Q,f}}
    &
    \malg{(\malg{Q} \comp f)}\, (\eta_A\, a)
    \ar[d]^-{\i_{\malg{Q} \comp f}}
    \\
    &
    \malg{Q}\, (f\, a)
  }
  \displaybreak[0]\\
  %
  \xymatrix@C=3em@R=3em@M=0.5em@L=0.5em{
    \malg{P}\, (\bind{\eta_A}\, m)
    \ar@{=}[dr]_-{\text{mon.~law}~~}
    \ar[r]^-{\j_{P,\eta_A}}
    &
    \malg{(\malg{P} \comp \eta_A)}\, m
    \ar[d]^-{\malgfun{\i_P}}
    \\
    &
    \malg{P}\, m
  }
 \displaybreak[1]\\
  \xymatrix@C=2.7em@R=1em@M=0.5em@L=0.5em{
    \malg{R}\, (\bind{(\bind{g} \comp f)}\, m)
    \ar@{=}[dd]_-{\text{mon.~law}}
    \ar[r]^-{\j_{R, \bind{g} \comp f}}
    &
    \malg{(\malg{R} \comp (\bind{g} \comp f))}\, m
    \ar[dr]^(0.6){\qquad\malgfun{\ilam{a} \j_{R,g} \, \ia{f\, a}}}
    \\
    &&
    {\malg{(\malg{(\malg{R} \comp g)} \comp f)}\, m}
    \\
    \malg{R}\, (\bind{g}\, (\bind{f}\, m))
    \ar[r]_-{\j_{R,g}}
    &
    \malg{(\malg{R} \comp g)}\, (\bind{f}\, m)
    \ar[ur]_-{\j_{\malg{R} \comp g , f}}
  }
\end{gather*}
\end{itemize}
The ingredients in the above diagrams are typed as follows:
$P,P' : A \to \U$,
$Q,Q' : B \to \U$,
$R : C \to \U$,
$h : \iprd{a \of A} (P\, a \to P'\, a)$,
$k : \iprd{b \of B} (Q\, b \to Q'\, b)$,
$f : A \to M B$, $g : B \to M C$, $a : A$, and $m : M A$.
\end{defn}

A mathematically pleasing stricter version would result if we required the maps~$\i_P$ and $\j_{P,f}$ to be families of type equivalences.
In fact, \Cref{def:weak-mendler-style-T-algebra} can be seen as a weakening of the notion of algebras for 2-categorical pseudomonads in no-iteration form~\cite{Marmolejo:NoIterationPseudomonads}, which requires the mediating 2-cells~$\i_P$ and~$\j_{Q,f}$ to be invertible.
However, such invertibility does not occur in all of our examples, such as in \cref{sec:self-represented-functionals}, and it is unnecessary in the general construction of monads on containers, as given in \cref{thm:monad-from-mendler-style-algebra} below.

The definition works with various notions of universes (Tarski or Russell-style, cumulative or not). Also note that by replacing~$\U$ with~$\Type$ (which itself is not a type), we obtain a notion of \emph{large} algebra acting on all types and families.

We avoid steering into even deeper categorical waters that would arise if we were to replace $\Type$ and~$\U$ with general categories~$\C$ and~$\D$, yielding a notion of weak Mendler-style $M$-algebra for a monad~$M$ on~$\C$.

The next theorem uses weak Mendler-style $M$-algebras to give us a general recipe for building monads on containers in the style of \cref{sec:finite-support}.
We write $\Cont(\U)$ for the category of containers whose positions $P : A \to \U$ are valued in~$\U$. The category $\Cont$ from \cref{sec:containers-morphisms} can be written as $\Cont(\Type)$ so long as we keep in mind that $\Type$ is not a type and that the result is a large category.

\begin{thm}
  \label{thm:monad-from-mendler-style-algebra}
  Given a monad $(M,\eta^M,\bind{(-)}_M)$ on $\Type$, a weak Mendler-style $M$-algebra on a universe~$\U$
  induces a monad $(T,\eta,\bind{(-)})$ on $\Cont(\U)$, given by
  \begin{equation*}
    \begin{array}{rcl}
      T(\cont A P) &\defeq& \cont {M A} {\malg{P}},
      \\
      \eta_{\cont A P} &\defeq& \cont {\eta^M_A} {(\ilam{a} \i_P\, \ia{a})},
      \\[0.5ex]
      \bind{(\cont f g)} &\defeq&
      \cont
        {\bind{f}_M}
        {\left(\ilam{m} (\malgfun{g}\, \comp \j_{Q,f}\, \ia{m})\right)}.
    \end{array}
  \end{equation*}
\end{thm}

\begin{proof}
  We need to establish the monad laws for $(T,\eta,\bind{(-)})$.
  On shapes they reduce to the laws for $(M,\eta^M,\bind{(-)}_M)$,
  so we only need to check them for positions.

  The first unit law $\bind {\eta_{\cont A P}} = \id_{T(\cont A P)}$ amounts to the coherence of~$\i$, as follows:
  \[
    \xymatrix@C=3em@R=3em@M=0.5em@L=0.5em{
      \malg{P}\, (\bind{\eta_A}\, m)
      \ar[r]^-{\j_{P,\eta_A}}
      \ar@{=}[d]_-{\text{mon.~law}}
      &
      \malg{(\malg{P} \comp \eta_A)}\, m
      \ar[r]^-{\malgfun{\i_P}}
      &
      \malg{P}\, m
      \ar@{=}[d]
      \\
      \malg{P}\, m
      \ar[rr]_-{\id_{\malg{P} m}}
      \ar@{}@<1.5em>[rr]^-{\dcomment{coherence of $\i$}}
      &
      &
      \malg{P}\, m
    }
  \]
  The second unit law
  $\bind {(\cont f g)} \comp \eta_{\cont A P} = \cont f g$ 
  amounts to commutativity of the following diagram:
  \[
    \xymatrix@C=3em@R=2em@M=0.5em@L=0.5em{
      \malg{Q}\, (\bind{f}\, (\eta_A\, a))
      \ar[r]^-{\j_{Q,f}}
      \ar@{=}@/_1pc/[dr]_-{\text{mon.~law}~~}
      \ar@{=}[dd]_-{\text{mon.~law}}
      &
      \malg{(\malg{Q} \comp f)}\, (\eta_A\, a)
      \ar[r]^-{\malgfun{g}}
      \ar[d]^-{\i_{\malg{Q} \comp f}}
      \ar@{}@<-0.3em>[d]_<<<<{\dcomment{coherence of $\i$}}
      \ar@{}@<3em>[d]^<<<<{\dcomment{naturality of $\i$}}
      &
      \malg{P}\, (\eta_A\, a)
      \ar[r]^-{\i_P}
      &
      P\, a
      \ar@{=}[dd]
      \\
      &
      \malg{Q}\, (f\, a)
      \ar@/_1pc/[urr]_-{g}
      &
      \\
      \malg{Q}\, (f\, a)
      \ar[rrr]_-{g}
      \ar@{}@<1em>[rrr]^-{\dcomment{=}}
      &
      &
      &
      P\, a
    }
  \]
  Finally, the associativity law $\bind {(\bind {(\cont k l)} \comp
  {\cont f g})} = \bind {(\cont k l)} \comp \bind {(\cont f g)}$ amounts to commutativity of the following diagram:
  \[
    \xymatrix@C=0.65em@R=3em@M=0.5em@L=0.5em{
      \malg{R}\, (\bind{(\bind{k} \comp f)}\, m)
      \ar@{=}[dd]_-{\text{mon.~law}}
      \ar[r]^-{\j_{R, \bind{k} \comp f}}
      \ar@{}@<1em>[dd]^>>>>>>>>>>>{\dcomment{coherence of $\j$}}
      &
      \malg{(\malg{R} \comp (\bind{k} \comp f))}\, m
      \ar[rrr]^-{\malgfun{\ilam{a} g \, \comp \, 
        \malgfun{l} \, \comp \, \j_{R,k}\, \ia{f\, a}}}
      \ar@{}@<-0.75em>[rrr]_(0.4){\dcomment{functoriality of $\malgfun{-}$}}
      \ar[d]_<<<<<<<<{\malgfun{\ilam{a} \j_{R,k}\, \ia{f\,a}}}
      &&&
      \malg{P}\, m
      \ar@{=}[dd]
      \ar@{}@<-3em>[dd]_>>>>>>>{\dcomment{=}}
      \\
      &
      \malg{(\malg{(\malg{R} \comp k)} \comp f)}\, m
      \ar[rr]^-{\malgfun{\ilam{a} \malgfun{l}\, \ia{f\, a}}}
      &
      &
      \malg{(\malg{Q} \comp f)}\, m
      \ar[ur]_-{\malgfun{g}}
      \\
      \malg{R}\, (\malg{k}\, (\malg{f}\, m))
      \ar[r]_-{\j_{R,k}}
      &
      \malg{(\malg{R} \comp k)}\, (\bind{f}\, m)
      \ar[r]_-{\malgfun{l}}
      \ar[u]^(0.4){\j_m^{\malg{R} \comp k , f}}
      &
      \malg{Q}\, (\bind{f}\, m)
      \ar[r]_-{\j_{Q,f}}
      \ar[ur]_-{\j_{Q,f}}
      \ar@{}@<1.5em>[ur]^(0.1){\dcomment{naturality of $\j$}}
      &
      \malg{(\malg{Q} \comp f)}\, m
      \ar[r]_-{\malgfun{g}}
      &
      \malg{P}\, m
    }\qedhere
  \]
\end{proof}

To familiarise ourselves with this construction, let us see how the finitely-supported functionals arise as an instance. As in \cref{sec:finite-support}, the monad~$M$ is the finite powerset monad $\FPow{}$ and
the extension operation takes families $P : A \to \Type$ to $\malg{P} : \FPow{A} \to \Type$ as $\malg{P}\, S
\defeq \prd{a \of S} P\, a$.
The functorial action takes maps of families $h : \iprd{a \of A} (P a \to Q a)$ to
$\malgfun{h} : \iprd{S \of \FPow A} (\malg{P} S \to \malg{Q} S)$, given by
\[
  \malgfun{h} \defeq \ilam{S} \lam{(k \of \prd{a
  \of S} P\, a)} \lam{(a \of S)} h\, \ia{a}\, (k\, a),
\]
Finally, the maps
\[
  \i_P : \iprd{a \of A} (\malg P\,\finset a \to P\, a)
  \quad\text{and}\quad
  \j_{Q,f} : \iprd{S \of \FPow A} (\malg Q\, (\bind f\, S) \to \malg {(\malg Q \comp f)}\, S)
\]
witnessing the lax Mendler-style algebra laws are given by
\[
  \begin{array}{l c l}
    \i_P & \defeq & \ilam{a} \lam{(h : \prd{a' \of \finset{a}} P\, a')} h\, a, 
    \\[3pt]
    \j_{Q,f} & \defeq & \ilam{S} \lam{(h : \prd{b \of \bind f S} Q\, b)} \lam{(a \of S)} h{\restriction}_{f\, a}, 
  \end{array}
\]
where $P : A \to \Type$, $Q : B \to \Type$, and $f : A \to \FPow B$.
A short calculation shows that applying \cref{thm:monad-from-mendler-style-algebra}
to this data does in fact result in the monad we defined in \cref{sec:finite-support}.

Note that the weak Mendler-style algebras produce monads that are unlike the monad of trees from
\cref{sec:comodule-representation-of-functionals}, because the shape part is always independent of the position part of the monad.

\subsection{Self-Represented Functionals}
\label{sec:self-represented-functionals}

We may apply \cref{thm:monad-from-mendler-style-algebra} to the trivial monad
$M X \defeq \One$, with the weak Mendler-style $M$-algebra extension operation
defined as $\malg P\, \star \defeq \prd{a \of A} P\, a$, for $P : A \to \Type$.
It is not hard to deduce the rest of the structure, so we only point out that
the map $\i_P : \iprd{a \of A} ((\prd{a' \of A} P\, a') \to P\, a)$, which is defined as
$\i_P\, \ia{a}\, h \defeq h\, a$, is not an isomorphism in general,
so we really do need the weak version of Mendler-style algebras.
Note that, as Richard Garner pointed out to us, the monad $T$ on $\Cont$ so
obtained also arises as the composition of $\cosem{-} : \Cont \to \Type^\op$ with its
right adjoint $A \mapsto \cont \One {(\lam{\_} A)}$.
We use the $T$-comodule structure map whose component
$\cook_{\cont A P} : \cosem {\cont A P} \to \cosem{T(\cont A P)}$
is defined as $\cook_{\cont A P}\, h\, \star \defeq h$.

The resulting notion of representation is trivial:
$F : \left(\prd{a \of A} P\, a\right) \to \left(\prd{b \of B} Q\, b\right)$
is represented by maps $\tree_F : B \to \One$ and
$\eat_F : \iprd{b \of B} ((\prd{a \of A} P\, a) \to Q\, b)$,
such that $F\, h\, b = \eat_F\, {\ia b}\, h$.
The map $\tree_F$ is trivial and $\eat_F$ is just~$F$ with swapped arguments,
which means that any $F$ is essentially represented by itself this way.

\subsection{Functional Functionals}
\label{sec:functional-functionals}

Another trivial, but slightly more interesting example arises when we take the identity monad $T(\cont A P) = \cont A P$ and the identity comodule structure map $\id : \cosem{{-}} \to \cosem{T({-})}$. This too is an instance of \cref{thm:monad-from-mendler-style-algebra}, for the identity monad and identity Mendler-style algebra.
Now a functional $F : \left(\prd{a \of A} P\, a\right) \to \left(\prd{b
\of B} Q\, b\right)$ is represented by maps $\tree_F : B \to A$ and $\eat_F :
\iprd{b \of B} (P\, (\tree_F\, b) \to Q\, b)$, such that $F\,h\,b = \eat_F (h (\tree_F\,b))$.
That is, $F\,h$ computes the answer at~$b$ by making a single query $\tree_F\,b$ to its argument~$h$.
We call it a \emph{functional} functional.

\subsection{Exceptional Functionals}
\label{sec:exceptional-functionals}

In the previous example, a represented functional made exactly one query to its input, even if the result did not depend on it.
By employing the exception monad $T(\cont A P) \defeq \cSum {(\cont A P)} {\cId}$, we may model functionals that either make one query or explicitly ignore the input.
We are still in the territory of \cref{thm:monad-from-mendler-style-algebra}, with $M\,A \defeq A + \One$ and the Mendler-style algebra extension operation  defined as
$\malg{P}(\inl\,a) \;\defeq\; P\,a$
and
$\malg{P}(\inr\,\star)  \;\defeq\; \One$, for all $P : A \to \Type$, and where ${\malg{P} : A + \One \to \Type}$.
Next, the comodule structure map at $\cont A P$ is $\cook_{\cont A P} : \cosem{\cont A P} \to \cosem{\cont {A + \One} {\malg P}}$, defined as
$\cook_{\cont A P} \, h \, (\inl\, a) \defeq h\, a$
and
$\cook_{\cont A P} \, h \, (\inr\, \star) \defeq \star$.

A functional $F : (\prd{a \of A} P\, a) \to (\prd{b
\of B} Q\, b)$ is represented in this sense when it takes the following form,
for some maps $\tree_F : B \to A + \One$ and
$\eat_F = [e, q] : \iprd{b \of B} (\malg P (\tree_F\, b) \to Q\,b)$, where
$e : \iprd{b \of B} P\, a \to Q\, b$ and $q : \iprd{b \of B} Q\, b$:
\begin{equation*}
  F\,h\,b =
  \begin{cases}
  e {\ia b}\, (h \, a),     &\text{if $\tree_F\,b = \inl\,a$,} \\
  q {\ia b},                &\text{if $\tree_F\,b = \inr\,\star$.}
  \end{cases}
\end{equation*}
Indeed, $\tree_F\,b$ either issues a query $a : A$, or signals that none is needed, after which the result is either computed from~$b$ and $h\,a$, or a default value $q\,{\ia b}$ is returned.


\section{Effectful Functionals}
\label{sec:effectful-functionals}

Taking inspiration from the exceptional functionals, in this section we study representations of functionals involving computational effects, expressed in terms of monads~\cite{Moggi:NotionsofComputationandMonads,Plotkin:NotionsOfComputations}.
We will be lead to rethink and vary the central \cref{def:comodule-representation} so that it incorporates effectful computations, see \cref{def:comodule-representation-parameterised-with-a-functor} below.

For concreteness, we work with the input-output effect,
even though everything generalises straightforwardly to any computational effect having an algebraic presentation~\cite{Plotkin:NotionsOfComputations} with a corresponding runner~\cite{ahman20:_runner_action,Uustalu:Runners,Plotkin:TensorsOfModels,Mogelberg:LinearState}, as we explain below.

Recall that the \emph{input-output monad} has its functor part $\IO{}$ defined
inductively by the following clauses, where~$I$ and~$O$ are fixed types of inputs and outputs:
\begin{mathpar}
  \inferrule
    {a : A}
    {\return(a) : \IO A}
  
  \inferrule
    {c : I \to \IO A}
    {\inp(c) : \IO A}

  \inferrule
    {o : O \\ c : \IO A}
    {\out(o, c) : \IO A}
\end{mathpar}
An element of $\IO A$ is an inductively generated tree representing a terminating effectful computation that may read $I$-valued inputs and write $O$-valued outputs by performing the operations $\inp$ and $\out$, and eventually return a value of type~$A$.
The constructor $\return$ serves as the unit of the monad, and the Kleisli extension grafts trees, i.e., given a map $f : A \to \IO B$, the map $\bind f : \IO A \to \IO B$ is defined as
\[
  \begin{array}{l c l}
    \bind f\, (\return(a)) & \defeq & f\, a, 
    \\[3pt]
    \bind f\, (\inp(c)) & \defeq & \inp(\lam{i} \bind f\, (c\, i)), 
    \\[3pt]
    \bind f\, (\out(o, c)) & \defeq & \out(o,\bind f\, c).
  \end{array}
\]

The interaction of $\IO$-computations with an environment is quite naturally modelled
using runners~\cite{ahman20:_runner_action,Uustalu:Runners}. Specifically, a \emph{stateful
$\IO$-runner} $\R = (R , \coinp{\R}, \coout{\R})$ consists of a type~$R$ of states (of the environment) together with \emph{co-operations}
$\coinp{\R} : R \to R \times I$ and $\coout{\R} : R \times O \to R$, modelling the behaviour of the operations~$\inp$ and~$\out$.
The runner induces a map $\rho_{\R,A} : \IO A \to R \to R \times A$, which \emph{runs} a computation in a given $R$-state to give a new state and a result:
\[
  \begin{array}{l c l}
    \rho_{\R,A} (\return(a))\, r & \defeq & (r,a), \\[3pt]
    \rho_{\R,A} (\inp(c))\, r & \defeq & \mathsf{let}\; (r',i) = \coinp{\R}\, r \; \mathsf{in}\; \rho_{\R, A} \, (c\, i)\, r', \\[3pt]
    \rho_{\R,A} (\out(o, c))\, r & \defeq & \rho_{\R,A}\, c\, (\coout{\R}\, (r,o)).
  \end{array}
\]
Speaking mathematically, $\rho_\R : \IO \to \St_R$ is a monad morphism from~$\IO$ to the $R$-valued \emph{state monad} $\St_R\, A \defeq R \to R \times A$ \cite{Moggi:NotionsofComputationandMonads}.
Moreover, $\rho_{\R,A}$ is the unique $\IO$-algebra morphism from the free algebra to the $\IO$-algebra on $\St_R\, A$ induced by the runner~$\R$, see~\cite{Mogelberg:LinearState}.

The last observation explains how the case of $\IO$ monad can be generalised to
monads whose algebraic presentations involve
equations~\cite{Plotkin:NotionsOfComputations,HylandPP:SumAndTensor}, such as
the monads for reading, writing, and state: so long as the runner $\R$ and the induced
algebra structure on $\St_R\, A$ satisfy the equations governing the monad, the universal property of the free algebra ensures the existence of~$\rho_R$.

There are several ways of combining computational effects and comodule representations of functionals, of which we shall consider three in this paper:
in \cref{sec:functional-effectful-functionals-second-attempt}, we work out a basic variant that uses effectful functional functionals with effectful arguments,
in \cref{sec:continuous-effectful-functionals}, we tackle effectful tree-represented functionals with effectful arguments,
and finally in \cref{sec:effectful-functionals-with-effect-free-representations}, we show that pure tree-represented functionals  with effectful arguments are a possibility, too.

\subsection{Effectful Functional Functionals With Effectful Arguments}
\label{sec:functional-effectful-functionals-second-attempt}

We wish to study second-order functionals that interact with
their environment. We expect the representable ones to be those that
interact through some well-behaved interface that limits access to the environment. Thus,
the so-called ``snap back'' functional, which resets the environment to a past one, should not be so represented.
For concreteness, we again restrict to the $\IO$ monad and take as the interface the input and output co-operations provided by the $\IO$-runner.

A natural first idea would be to proceed just like in the case of exceptional functionals from \cref{sec:exceptional-functionals}, but with the~$\IO$ monad in place of the exception monad. If \Cref{thm:monad-from-mendler-style-algebra} can still be made to work, and a comodule structure map is found, a notion of ``$\IO$-represented'' functionals will be obtained.
One can get quite far in defining the required Mendler-style $\IO$-algebra structure, with respect to a given $\IO$-runner $\R$. Specifically, given a type family $P : A \to \Type$, we can define $\malg P : \IO A \to \Type$ as $\malg P c \defeq P\, (\snd\,
(\rho_{R,A}\, c\, r_\text{init}))$, where $r_\text{init} : R$ is some fixed inital runner state.
Thus $\malg P c$ runs the computation~$c$ in the initial state, discards the final state, and passes the computed result to~$P$.
However, the associativity law for $\malg{({-})}$ fails because the runner state keeps getting reset to the chosen initial one~$r_\text{init}$. Instead, we need a version that correctly propagates the state of the runner through the computation of the functional.

To correctly account for state propagation, we instead consider functionals
\[
  F : (\prd{a \of A} (R \to R \times P\, a)) \to (\prd{b \of B} (R \to R \times Q\, b)),
\]
where $R$ is the carrier of an $\IO$-runner $\R = (R, \coinp{\R},
\coout{\R})$. 
The argument of~$F$ and its result both have access to the runner state, which allows them to perform $\IO$-operations, as well as to manipulate the state in any way they see fit. However, we expect~$F$ to have a comodule representation only when it interfaces with the state strictly via the $\IO$-operations.

We must adapt \cref{def:comodule-representation} to the altered form of functionals. Let us do so in general, by considering any functor $S : \Type \to \Type$ and functionals 
\[
  F : (\prd{a \of A} S (P\, a)) \to (\prd{b \of B} S (Q\, b)).
\]
We write $\cosem{-}_S$ for the composition of $\cosem{-}$ with $S$ acting on positions, given by
\[
  \begin{array}{l c l}
    \cosem{\cont A P}_S & \defeq & \prd{a \of A} S\, (P\, a), \\[3pt]
    \cosem{\cont f g}_S\, h & \defeq & \lam{(a \of A)} S(g)\, (h\, (f\, a)),
  \end{array}
\]
so that we can write more succinctly $F : \cosem{\cont A P}_S \to \cosem{\cont B Q}_S$.
The stateful variant we started with is recovered when we instantiate~$S$ with the state monad~$\St_R$.

\begin{defn}
  \label{def:comodule-representation-parameterised-with-a-functor}
  Let $(T,\eta,\bind{(-)})$ be a monad on $\Cont$, $S : \Type \to \Type$ a
  functor, and ${\cook : \cosem{-}_S \to \cosem{T(-)}}_S$ a right $T$-comodule
  structure map on $\cosem{{-}}_S$.
  A \emph{$(T,\cosem{-}_S,\cook)$-representation} of a second-order functional 
  \[ 
    F : \left(\prd{a \of A} S\, (P\, a)\right) \to \left(\prd{b \of B} S\, (Q\, b)\right)
  \] 
  is a container morphism 
  $\cont {\tree_F}
  {\eat_F} : \cont B Q \to T(\cont A P)$ satisfying
  \begin{equation}
    \label{dia:comodule-representation-parameterised-with-a-functor}
    \begin{gathered}
    \xymatrix@C=1em@R=3em@M=0.5em@L=0.5em{
      \cosem{\cont A P}_S
      \ar[rr]^-{F}
      \ar@/_1pc/[dr]_(0.25){\cook_{\cont A P}}
      &&
      \cosem{\cont B Q}_S
      \\
      &
      \cosem{T(\cont{A}{P})}_S
      \ar@/_1pc/[ur]_(0.75){\cosem{\cont {\tree_F} {\eat_F}}_S}
    }
    \end{gathered}
  \end{equation}
\end{defn}

\noindent
\Cref{def:comodule-representation} is an instance of \cref{def:comodule-representation-parameterised-with-a-functor} arising from $S \defeq \Id$.

The functionals so represented constitute a category $\RFun{T,\cosem{-}_S,\cook}$ in analogy with \cref{thm:comodule-representable-functionals-category}.
Notice how we made sure that a represented functional interacts with~$S$ only through the monad~$T$ by placing~$S$ in the cointerpretation in the domain and codomain of the functional, as opposed to indexing the comodule structure map with containers of the form $\cont A {(S \circ P)}$.
Specifically, $\IO$-representable functionals will interact with the runner state~$R$ solely through $\IO$-operations.

To complete the construction of $\IO$-representations of functionals, now written as
$F : \cosem{\cont A P}_{\St_\R} \to \cosem{\cont B Q}_{\St_\R}$,
we still need a monad~$T$ and a right $T$-comodule to go with it. We obtain the former from \cref{thm:monad-from-mendler-style-algebra} with $M \defeq \IO$ and a weak Mendler-style $\IO$-algebra structure that we define next.
Given a container $\cont A P$ and $c : \IO A$, let the \emph{traces} $\IOTrace[A,P]{c}$ be generated inductively by
\begin{mathpar}
  \inferrule
    {p : P\, a}
    {\iostop(p) : \IOTrace[A,P]{\return(a)}}

  \inferrule
    {i : I \\ \trc : \IOTrace[A,P]{c\, i}}
    {\istep(i, \trc) : \IOTrace[A,P]{\inp(c)}}

  \inferrule
    {\trc : \IOTrace[A,P]{c}}
    {\ostep(\trc) : \IOTrace[A,P]{\out(o, c)}}
\end{mathpar}
A trace $\tau : \IOTrace[A,P]{c}$ can be thought of as a finite sequence recording the inputs read and outputs written by an execution of a computation~$c$ terminating with a query $a : A$, with the trace correspondingly stopping with an answer $p : P\,a$. (The actual execution of a computation is carried out by the comodule structure map, as explained below.)
Note that the constructor~$\ostep$ does not record the output value, but just the fact that output happens, as the value can be read off the computation.

The weak Mendler-style extension $\malg P c \defeq \IOTrace[A,P]{c}$ carries the required structure. Given $h : \iprd{a \of A} P\, a \to Q\, a$, the functorial action is defined as follows:
\[
  \begin{array}{l c l}
    \multicolumn{3}{l}{\malgfun{h} : \iprd{c \of \IO A} \IOTrace[A,P]{c} \to \IOTrace[A,Q]{c},} 
    \\[3pt]
    \malgfun{h}\, \ia{\return\, a}\, (\iostop(p)) & \defeq & \iostop(h\, p), 
    \\[3pt]
    \malgfun{h}\, \ia{\inp(c)}\, (\istep(i,\trc)) & \defeq & \istep(i,\malgfun{h}\, \ia{c\, i}\, \trc), 
    \\[3pt]
    \malgfun{h}\, \ia{\out(o,c)}\, (\ostep(\trc)) & \defeq & \ostep(\malgfun{h}\, \ia{c}\, \trc), 
  \end{array}
\]
the map laxly witnessing the first Mendler-style $\IO$–algebra law is defined as
\[
  \begin{array}{l c l}
    \multicolumn{3}{l}{\i_P : \iprd{a \of A} \IOTrace[A,P]{\return\, a} \to P\, a,}
    \\[3pt]
    \i_P\, (\iostop(p)) & \defeq & p, 
  \end{array}
\]
and the map laxly witnessing the second Mendler-style $\IO$–algebra law is given by
\[
  \begin{array}{l c l}
    \multicolumn{3}{l}{\j_{Q,f} : \iprd{c \of \IO A} \IOTrace[B,Q]{\bind f\, c} \to \IOTrace[{A,\lam{a} \IOTrace[B,Q]{f\, a}}]{c}, }
    \\[3pt]
    \j_{Q,f}\, \ia{\return\, a}\, \trc & \defeq & \iostop(\trc), 
    \\[3pt]
    \j_{Q,f}\, \ia{\inp(c)}\, (\istep(i,\trc)) & \defeq & \istep(i,\j_{Q,f}\, \ia{c\, i}\, \trc), 
    \\[3pt]
    \j_{Q,f}\, \ia{\out(o,c)}\, (\ostep(\trc)) & \defeq & \ostep(\j_{Q,f}\, \ia{c}\, \trc).
  \end{array}
\]
As a result, \cref{thm:monad-from-mendler-style-algebra} yields a monad $T(\cont
A P) = \cont {\IO A} {(\lam{c} \IOTrace[A,P]{c})}$.

It remains to define a right $T$-comodule structure map on $\cosem{-}_{\St_R}$. The component $\cook_{\cont A P} : \cosem{\cont A P}_{\St} \to
\cosem {T(\cont A P)}_{\St}$ at a container $\cont A P$ is the map
\[
  \cook_{\cont A P} : (\prd{a \of A} (R \to R \times P\, a)) \to 
    (\prd{c \of \IO A} (R \to R \times \IOTrace[A,P]{c})),
\]
defined recursively, with the aid of the available $\IO$-runner $\R$, as follows:
\[
\begin{array}{l c l}
  \cook_{\cont A P}\, h\, (\return(a))\, r & \defeq & \mathsf{let}\; (r',p) = h\, a\, r\; \mathsf{in}\; (r',\iostop(p)),  \\[3pt]
  \cook_{\cont A P}\, h\, (\inp(c))\, r & \defeq & \mathsf{let}\; (r',i) = \coinp{\R}\, r\; \mathsf{in}\; \\[1pt]
    && \mathsf{let}\; (r'',\trc) = \cook_{\cont A P}\, h\, (c\, i)\, r'\; \mathsf{in}\; (r'', \istep(i,\trc)),  \\[3pt]
  \cook_{\cont A P}\, h\, (\out(o, c))\, r & \defeq & \mathsf{let}\; r' = \coout{\R}\, (r,o)\; \mathsf{in}\; \\[1pt]
    && \mathsf{let}\; (r'',\trc) = \cook_{\cont A P}\, h\, c\, r'\; \mathsf{in}\; (r'',\ostep(\trc)).
\end{array}
\]
The map just runs a computation using the runner~$\R$ in an initial state~$r$, and composes a trace of inputs and outputs. When a value is returned, it consults~$h$ to obtain an answer that is also recorded in the trace. The state is threaded through the computation, as well as through the potentially state-manipulating argument~$h$,
to ensure compositional behaviour that validates the required comodule laws.

According to \cref{def:comodule-representation-parameterised-with-a-functor}, a
functional $F : \cosem{\cont A P}_{\St_\R} \to \cosem{\cont B Q}_{\St_\R}$ represented
by maps $\tree_F : B \to \IO A$ and $\eat_F : \iprd{b \of B} (\IOTrace[A,P]{\tree_F\, b} \to Q\, b)$
computes as follows. Given $r : R$, $h : \cosem{\cont A P}_{\St_\R}$, and $b : B$,
the $\IO$-computation $\tree_F\,b$ is executed in state~$r$. A trace of the execution, which includes the answer 
by~$h$ to a single query, is passed to~$\eat_F$, which gives a result in $Q\,b$.
As noted earlier, the state is threaded through all the way.
We call such a functional \emph{interactive}.

The algebra representations from \cref{sec:algebra-representations-of-functionals}
allow us to arrive at the above rather concrete definitions by abstract means.
First, $T(\cont A P)$ is isomorphic to the functor part of the free monad on
$\Cont$ for the $\IO$ operations' signature, which is constructed as the initial algebra
of the endofunctor on $\Cont$ given by
\[
  \cont B Q
  \ \mapsto \ 
  \cSum {(\cont A P)} {
  \cSum {(\cont B Q)^I} 
        {O \cdot (\cont B Q)}}.
\]
Here $(\cont B Q)^I$ is the $I$-th power and $O \cdot (\cont B Q)$ the $O$-fold copower of $\cont B Q$:
\begin{align*}
  (\cont B Q)^I &\defeq \cExp{(\cont I {(\lam{\_} \Zero)})}{(\cont B Q)}, \\
  O \cdot (\cont B Q) &\defeq \cProd{(\cont O {(\lam{\_} \Zero)})}{(\cont B Q)}.
\end{align*}
Next, the functor $\cosem{-}_{\St_\R}$ is represented by the container $(R \cdot \cId)^R$.
Indeed, as $(R \cdot \cId)^R \cong \cont {(R
\to R)} {(\lam{\_} R)}$, a container map $\cont A P  \to (R \cdot \cId)^R$
is given by maps $A \to R \to R$ and $\prd{\ia{a \of A}} (R \to P\, a)$, which is
the same as giving a combined map $\prd{a \of A} (R \to R \times P\, a)$.
Finally, as $R$ is the carrier of an $\IO$-runner, $(R
\cdot \cId)^R$ naturally forms an $\IO$-algebra in $\Cont$, and thus
carries a $T$-algebra structure, from which the comodule structure map~$\cook$ is obtained as in \cref{sec:algebra-representations-of-functionals}.

The algebra representation just given can be modified to give variations on a theme.
We can replace the state monad~$\St_\R$ with the $R$-valued reader monad $\Rd_\R\, X \defeq R
\to X$. The resulting functor $\cosem{\cont A P}_{\Rd_\R} = \prd {a \of A} (R \to P\, a)$ is represented by the container $(\cId)^R$, which also carries an $\IO$-algebra in~$\Cont$. A corresponding notion of represented functionals $\cosem{\cont A P}_{\Rd_\R} \to \cosem{\cont B Q}_{\Rd_\R}$ arises.
We could also generalise to the (not a monad) functor $J X \defeq R \to R' \times X$,
with potentially different input and output states $R$ and $R'$. The combined functor $\cosem{-}_J$ is represented by the container $(R' \cdot \cId)^R$, which also forms an $\IO$-algebra.

\subsection{Effectful Tree-Represented Functionals With Effectful Arguments}
\label{sec:continuous-effectful-functionals}

The effectful behaviour of a functional from the previous section is quite simplistic, as it always performs a sequence of operations, after which it issues a single query to its argument.
In this section, we combine effects with tree-represented functionals that exhibit a more sophisticated computational behaviour.

Let us begin by constructing a monad on containers that captures the intuition that, in the course of its computation,
an interactive functional interleaves $\IO$ operations and queries to its argument in an arbitrary way.
This is done easily enough by simply combining the constructors for well-founded trees and input-output computations into a single inductive definition, and similarly for paths.
Concretely, we take the carrier to be the functor $T (\cont A P) \defeq
\cont {\IOTree{A,P}} {(\lam{t} \IOPath[A,P]{t})}$ on $\Cont$, where
$\IOTree{A,P}$ and $\IOPath[A,P]{t}$ are given inductively by the clauses
\begin{mathpar}
  \inferrule 
    {\phantom{\IOTree{A,P}}} 
    {\leaf : \IOTree{A,P}}
  
  \inferrule 
    {a : A \qquad \trs{t} : P\, a \to \IOTree{A,P}} 
    {\node(a,\trs{t}) : \IOTree{A,P}}
  \\
  
  \inferrule
    {\trs{t} : I \to \IOTree{A,P}}
    {\inp(\trs{t}) : \IOTree{A,P}}

  \inferrule
    {o : O \\ t : \IOTree{A,P}}
    {\out(o, t) : \IOTree{A,P}}
\end{mathpar}
and
\begin{mathpar}
  \inferrule
     {\phantom{\IOPath[A,P]{t}}} 
     {\pstop : \IOPath[A,P]{\leaf}}
    
  \inferrule 
     {p : P\, a \qquad \pi : \IOPath[A,P]{\trs{t}\, p}} 
     {\pstep(p,\pi) : \IOPath[A,P]{\node(a,\trs{t})}}

  \inferrule
     {i : I \\ \pi : \IOPath[A,P]{\trs{t}\, i}}
     {\istep(i, \pi) : \IOPath[A,P]{\inp(\trs{t})}}
 
   \inferrule
     {\pi : \IOPath[A,P]{t}}
     {\ostep(\pi) : \IOPath[A,P]{\out(o, t)}}
 \end{mathpar}
The corresponding unit and Kleisli extension are defined as in
\cref{sec:tree-representation}: the unit returns single-node trees, and the
Kleisli extension performs tree grafting.

Just like in the previous section, compositional treatment of effects is achieved by
considering functionals of the form $F : \cosem{\cont A P}_{\St_\R} \to \cosem{\cont B Q}_{\St_\R}$.
The component of the corresponding comodule structure map at $\cont A P$ has type
\[
  \cook_{\cont A P} : \left(\prd{a \of A} (R \to R \times P\, a)\right) \to 
    \left(\prd{t \of \IOTree{A,P}} (R \to R \times \IOPath[A,P]{t})\right).
\]
It combines the definitions of the comodule structure map
for tree-representations and $\IO$-computations, for a fixed runner $\R = (R, \coinp{\R}, \coout{\R})$, to produce combined paths of answers to queries and results of $\IO$ operations, as follows:
\[
\begin{array}{l c l}
  \cook_{\cont A P}\, h\, \leaf\, r & \defeq & (r , \pstop), \\[3pt]
  \cook_{\cont A P}\, h\, (\node(a,\trs{t}))\, r & \defeq & \mathsf{let}\; (r',p) = h\, a\, r \; \mathsf{in}\; \\[1pt]
    && \mathsf{let}\; (r'',\pi) = \cook_{\cont A P}\, h\, (\trs{t}\, p)\, r' \; \mathsf{in}\; (r'',\pstep(p,\pi)), \\[3pt]
  \cook_{\cont A P}\, h\, (\inp(\trs{t}))\, r & \defeq & \mathsf{let}\; (r',i) = \coinp{\R}\, r\; \mathsf{in}\; \\[1pt]
    && \mathsf{let}\; (r'',\pi) = \cook_{\cont A P}\, h\, (\trs{t}\, i)\, r'\; \mathsf{in}\; (r'', \istep(i,\pi)),  \\[3pt]
  \cook_{\cont A P}\, h\, (\out(o, t))\, r & \defeq & \mathsf{let}\; r' = \coout{\R}\, (r,o)\; \mathsf{in}\; \\[1pt]
    && \mathsf{let}\; (r'',\pi) = \cook_{\cont A P}\, h\, t\, r'\; \mathsf{in}\; (r'',\ostep(\pi)).
\end{array}
\]
The query to the argument~$h$ in the second clause is effectful, in contrast to the earlier comodule for pure (effect-free) tree-represented functionals.

In the present case, \cref{def:comodule-representation-parameterised-with-a-functor} tells us that a
functional 
\[
  F : \cosem{\cont A P}_{\St_\R} \to \cosem{\cont B Q}_{\St_\R}
\]
is represented by maps 
\[
  \tree_F : B \to \IOTree{A,P}
  \quad\text{and}\quad
  \eat_F : \iprd{b \of B} (\IOPath[A,P]{\tree_F\, b} \to Q\, b).
\]
These are used to compute $F\,h\,b$ in much the same way as before. The comodule structure map runs $\tree_F\,b$ in an initial state~$r$ to compute a combined trace, after which~$\eat_F$ computes the result; and the state is threaded through the entire computation.

Once again, the same notion of representability can be obtained using algebra representations from \cref{sec:algebra-representations-of-functionals}. The above monad~$T$
is isomorphic to the coproduct of the free monad from
the previous section and the tree monad from \cref{sec:tree-monad}. 
The representation $(R \cdot \cId)^R$ of $\cosem{-}_{\St_\R}$ carries the required algebra structure because it does so for each summand separately. Indeed, the one for the free monad was given in the previous section,
while an algebra for the tree monad results from recalling that it is the free monoid monad and noting that
$(R \cdot \cId)^R$ is evidently a monoid for~$\cComp{}{}$.

\subsection{Effectful Functionals With Effect-Free Representations}
\label{sec:effectful-functionals-with-effect-free-representations}

We conclude this discussion with a study of functionals that operate on effectful arguments, but are themselves pure, so that any computational effects must originate from the argument. This is a frequent situation in functional programming, for example when mapping or folding an effectful function over a list.

We limit attention to tree representations of functionals that operate on effectful functions
whose effects are embodied by a monad $(S,\eta^S,\bind {(-)}_S)$.
The fundamental representation diagram thus takes the following shape, where $\Tr$ is the tree monad from \cref{sec:tree-monad}:
\begin{equation*}
  \xymatrix@C=1em@R=3em@M=0.5em@L=0.5em{
    \cosem{\cont A P}_S
    \ar[rr]^-{F}
    \ar@/_1pc/[dr]_(0.25){\cook_{\cont A P, S}}
    &&
    \cosem{\cont B Q}_S
    \\
    &
    \cosem{\Tr(\cont{A}{P})}_S
    \ar@/_1pc/[ur]_(0.75){\cosem{\cont {\tree_F} {\eat_F}}_S}
  }
\end{equation*}
The only missing piece is the comodule structure map, which we define recursively:
\[
\begin{array}{l c l}
  \cook_{\cont A P, S}\, h\, \leaf & \defeq & \eta^S_{\Path[A,P]{\leaf}}\, \pstop,
  \\[3pt]
  \cook_{\cont A P, S}\, h\, (\node(a,t)) & \defeq &
  \bind{\big(\lam {p} \bind{(\lam {\pth{p}} \eta^S_{}\, (\pstep(p,\pth{p})))}\, (\cook_{\cont A P, S}\, h\, (t\, p))\big)} (h\, a).
\end{array}
\]
The first clause just embeds the trivial path into the monad $S$ using its unit.
The second clause is like the analogous one in the pure case from \cref{sec:tree-representation},
except that it takes into account the ambient monad $S$. It is easier understand when its right-hand side is written Haskell-style using $\mathsf{do}$-notation:
\begin{equation*}
  \mathsf{do}\;\{
   p \leftarrow h\,a ;\;
   \vec{p} \leftarrow \cook_{\cont A P, S} \, h\, (t\, p) ;\;
   \mathsf{return} (\pstep(p,\pth{p}))
  \}.
\end{equation*}

We now establish certain properties of the comodule map~$\cook$, which we shall use to make precise the claim that represented functionals are not the source of any effects.

\begin{prop}
  \label{prop:morphism-of-tree-comodules}
  Given a monad morphism $\theta : S \to S'$, the
  following diagram commutes for every $\cont A P$:
  \[
    \xymatrix@C=2em@R=2em@M=0.5em@L=0.5em{
      \cosem{\cont A P}_{S}
      \ar[r]^-{\cook_{\cont A P, S}}
      \ar[d]_-{\cosem{\cont A P}_{\theta}}
      &
      \cosem{\Tr(\cont A P)}_{S}
      \ar[d]^-{\cosem{\Tr(\cont A P)}_{\theta}}
      \\
      \cosem{\cont A P}_{S'}
      \ar[r]_-{\cook_{\cont A P, S'}}
      &
      \cosem{\Tr(\cont A P)}_{S'}
    }
  \]
  where $\cosem{\cont A P}_{\theta} : \cosem{\cont A P}_{S} \to \cosem{\cont A P}_{S'}$ is defined as $\cosem{\cont A P}_{\theta}\, h\, a \defeq \theta_{P\, a}\, (h\, a)$.
\end{prop}

\begin{proof}
  Showing that this diagram commutes amounts to proving that
  \[
    \cosem{\Tr(\cont A P)}_{\theta}\, (\cook_{\cont A P, S}\, h)\, t 
    =
    \cook_{\cont A P, S'}\, (\cosem{\cont A P}_{\theta}\, h)\, t, 
  \]
  for all $h : \cosem{\cont A P}_{S}$ and $t : \Tree{A,P}$. The proof proceeds
  by induction on the structure of $t : \Tree{A,P}$ and by straightforward
  diagram chasing, using both monad morphism laws of $\theta$.
\end{proof}

The above proposition says that a monad morphism
$\theta : S \to S'$ induces a morphism of right $\Tr$-comodules.
We now use this fact to show that there is also a corresponding morphism between
representable functionals:

\begin{prop}
  \label{prop:morphism-of-effectful-representable-functionals}
  Given a monad morphism $\theta : S \to S'$, a
  $(\Tr,\cosem{-}_S,\cook_{S})$-representable functional $F : \cosem{\cont A
  P}_S \to \cosem{\cont B Q}_S$ and a
  $(\Tr,\cosem{-}_{S'},\cook_{S'})$-representable functional $G : \cosem{\cont A
  P}_{S'} \to \cosem{\cont B Q}_{S'}$, such that $F$ and $G$ are represented by
  the same container morphism, then the next diagram
  commutes: 
  \[
    \xymatrix@C=2em@R=2em@M=0.5em@L=0.5em{
      \cosem{\cont A P}_S
      \ar[r]^-{F}
      \ar[d]_-{\cosem{\cont A P}_\theta}
      &
      \cosem{\cont B Q}_S
      \ar[d]^-{\cosem{\cont B Q}_\theta}
      \\
      \cosem{\cont A P}_{S'}
      \ar[r]_-{G}
      &
      \cosem{\cont B Q}_{S'}
    }
  \]
\end{prop}

\begin{proof}
  Let $\cont {\tree} {\eat}$ be the container morphism that represents both $F$
  and $G$. Consider
  \begin{equation}
    \label{eq:morphism-of-tree-comodules}
    \begin{gathered}
    \xymatrix@C=4em@R=2em@M=0.5em@L=0.5em{
      \cosem{\cont A P}_S
      \ar[rr]^-{F}
      \ar[ddd]_-{\cosem{\cont A P}_\theta}
      \ar[dr]_-{\cook_{\cont A P, S}}
      \ar@{}@<-0.5em>[rr]_-{\dcomment{$\cont{\tree}{\eat}$ represents $F$}}
      \ar@{}@<0.5em>[ddd]^-{\dcomment{\cref{prop:morphism-of-tree-comodules}}}
      &&
      \cosem{\cont B Q}_S
      \ar[ddd]^-{\cosem{\cont B Q}_\theta}
      \ar@{}@<-2.5em>[ddd]_-{\dcomment{naturality}}
      \\
      &
      \cosem{\Tr(\cont A P)}_S
      \ar[ur]_-{\cosem{\cont {\tree} {\eat}}_S}
      \ar[d]_-{\cosem{\Tr(\cont A P)}_\theta}
      \\
      &
      \cosem{\Tr(\cont A P)}_{S'}
      \ar[dr]^-{\cosem{\cont {\tree} {\eat}}_{S'}}
      \\
      \cosem{\cont A P}_{S'}
      \ar[rr]_-{G}
      \ar[ur]^-{\cook_{\cont A P, S'}}
      \ar@{}@<0.5em>[rr]^-{\dcomment{$\cont{\tree}{\eat}$ represents $G$}}
      &&
      \cosem{\cont B Q}_{S'}
    }
    \end{gathered}
  \end{equation}
\end{proof}

\noindent
There is nothing special about the role of the tree monad in the above statement, which could thus be generalised to an arbitrary monad~$T$ on $\Cont$ and a morphism of right $T$-comodules.

When we instantiate $S \defeq \Id$ and $S' \defeq M$, for some monad $(M,
\eta, \bind {(-)})$, and set $\theta \defeq \eta$, the diagram \eqref{eq:morphism-of-tree-comodules}
makes precise the intuition that all effects emanate from the arguments (we renamed~$F$ to $G_\text{pure}$ in the diagram to bolster this point):
\begin{equation}
  \label{eq:pure-to-effectful}
  \begin{gathered}
    \xymatrix@C=4em@R=2em@M=0.5em@L=0.5em{
      \cosem{\cont A P}
      \ar[rr]^-{G_{\text{pure}}}
      \ar[ddd]_-{\cosem{\cont A P}_{\eta}}
      \ar[dr]_-{\cook_{\cont A P}}
      &&
      \cosem{\cont B Q}
      \ar[ddd]^-{\cosem{\cont B Q}_{\eta}}
      \\
      &
      \cosem{\Tr(\cont A P)}
      \ar[ur]_-{\cosem{\cont {\tree_F} {\eat_F}}}
      \ar[d]_-{\cosem{\Tr(\cont A P)}_{\eta}}
      \\
      &
      \cosem{\Tr(\cont A P)}_{M}
      \ar[dr]^-{\cosem{\cont {\tree_F} {\eat_F}}_{M}}
      \\
      \cosem{\cont A P}_{M}
      \ar[rr]_-{G}
      \ar[ur]^-{\cook_{\cont A P, M}}
      &&
      \cosem{\cont B Q}_{M}
    }
  \end{gathered}
\end{equation}
Indeed, the inner part of the diagram expresses the fact that $G$ and $G_\text{pure}$ compute using the same dialogue tree, and the outer that restricting~$G$ to pure arguments gives a functional~$G_\text{pure}$ which factors through pure results, so it is pure.
In fact, \cref{prop:morphism-of-effectful-representable-functionals} says more than that: in the presence of two kinds of effects $S$ and $S'$, and a transformation~$\theta$ between them, there is a well-defined notion of represented ``$S$-pure'' functionals acting on ``$S'$-effectful'' arguments via~$\theta$.

What are some interesting choices of the monad $M$ in~\eqref{eq:pure-to-effectful}?
One option is a partiality or delay monad~\cite{Capretta:DelayMonad}, in which case computing $F\,
h\, b$ could be partial on the account of querying the argument~$h$ at an undefined value.
Another is the state monad $\St_\R$, as considered in~\cref{sec:continuous-effectful-functionals},
in which case any interaction of $F\, h\, b$ with the state is due to~$h$. 
As already mentioned, examples can be found in libraries for functional programming with container-like data structures, where iterators, maps, and folds apply (effectful) functions to data stored in a container.

In contrast to the last two sections, the $\Tr$-algebra
viewpoint of \cref{sec:algebra-representations-of-functionals} is not directly
applicable to the functor $\cosem{-}_S$ and the comodule defined above, at least not for
arbitrary~$S$. By following the methods of
\cref{sec:algebra-representations-of-functionals}, one notices that $\prd {a \of
A} S(P\, a) \cong \Cont(\Type_S)(\cont A P, \cId)$, where $\Cont(\Type_S)$ is
the category of containers whose position maps are Kleisli maps for~$S$. It stands
to reason that we should therefore consider monad algebras in $\Cont(\Type_S)$.
However, the tree monad $\Tr$ is not even a functor on this category. Naive
attempts at defining a tree monad on $\Cont(\Type_S)$ suggest combining trees and~$S$, but it is not clear how one might do that for an arbitrary~$S$. We leave further investigation as future work. 


\section{Instance Reductions}
\label{sec:propositional-containers}

We now steer in a different direction and explore proofs of implications between
universally quantified logical statements
\begin{equation}
  \label{eq:all-imply-all}
  (\all{a \of A} P\,a) \lthen (\all{b \of B} Q\,b).
\end{equation}
These arise in our setup when~$A,P$ and~$B,Q$ are taken to be \emph{propositional containers}, by which we mean
type families $P : A \to \Prop$ and $Q : B \to \Prop$ mapping to a universe~$\Prop$ of propositions that supports the usual connectives and quantifiers of intuitionistic logic. We write $\pcont A P$ to indicate that a container is propositional.
The cointerpretation restricts to $\cosem{-} : \PCont^\op \to \Prop$ as $\cosem{\pcont A P} = \all{a \of A} P\,a$.

In the constructions that follow, we assume that $\Prop$ satisfies the principles of \emph{proof irrelevance} and \emph{propositional extensionality}, respectively given by
\[
  \all{p \of \Prop} \all{x\,y \of p} x = y
  \qquad\text{and}\qquad
  \all{p\,q \of \Prop} (p \liff q) \lthen p = q.
\]
Examples include a subobject classifier in a topos~\cite{lambek-scott:88}, a type-theoretic impredicative universe of propositions~\cite{coquand88:_calcul_const,coquand85:_const} with the two principles additionally assumed to hold,
and a univalent type theory~\cite{hottbook} with resizing (needed to collapse the stratification of propositions by universe levels).
A pure propositions-as-types interpretation equating $\Prop$ with $\Type$ however cannot be used.

Propositional containers form a full subcategory~$\PCont$ of $\Cont$.
A morphism $\pcont f g : \pcont A P \to \pcont B Q$ consists of a map~$f : A \to B$ and a proof~$g$ of $\all{a \of A} Q(f\,a) \lthen P\,a$.
A more traditional way of writing this information is
\begin{equation*}
  \pcont A P \to \pcont B Q
  \;\defeq\;
  \{f : A \to B \mid \all{a \of A} Q(f\,a) \lthen P\,a\},
\end{equation*}
and we shall do so, by writing just $f : \pcont A P \to \pcont B Q$ in the rest of this section.

Next, we observe that a functional functional from \cref{sec:functional-functionals} restricted to~$\PCont$ amounts to
witnessing~\eqref{eq:all-imply-all} with a morphism $\tree : \pcont B Q \to \pcont A P$.
That is, \eqref{eq:all-imply-all} is proved in a particular way: assuming $\all{a \of A} P\,a$, for any $b : B$ we derive $Q\,b$ by applying modus ponens to $P(\tree\,b) \lthen Q\,b$ and $P(\tree\,b)$.
It is fitting to name such a representation a \emph{functional instance reduction}, as it is the explicit version of
\emph{functional instance reducibility} of $\pcont B Q$ to $\pcont A P$, defined as
\begin{equation}
  \label{eq:functional-instance-reducibility}%
  \some{t : B \to A} \all{b \of B} (P(t\,b) \lthen Q\,b).
\end{equation}
To put it another way, functional instance reducibility is the preorder reflection\footnote{Recall that the preorder reflection on a category $\C$ is the preorder~$\leq$ on the objects of~$\C$ in which $x \leq y$ holds when there exists a morphism $x \to y$.} of~$\PCont$.
In turn, \eqref{eq:functional-instance-reducibility} is a special case of general \emph{instance reducibility}~\cite{Bauer:InstanceReducibility}
\begin{equation}
  \label{eq:instance-reducibility}%
  \all{b \of B} \some{a \of A} (P\, a \lthen Q \, b),
\end{equation}
whose connection to comodule representations we address below.
For the moment, we continue to study functional instance reductions, as in mathematical practice they are the prevalent method of establishing statements of the form~\eqref{eq:all-imply-all}.

\begin{prop}
  \label{thm:pcont-distr-wexp}
  The category of propositional containers is distributive and has weak exponentials.
\end{prop}

\begin{proof}
  The finite sums and products are computed as in $\Cont$, see \cref{sec:containers-morphisms}, except that we use
  suitable constructs in~$\Prop$ instead of~$\Type$:
  \begin{align*}
    \pOne &\;\defeq\; \pcont \One {\lam{\_} \bot},
    \\
    \pZero &\;\defeq\; \pcont \Zero {\lam{\_} \bot}, 
    \\
    \pProd {(\pcont A P)} {(\pcont B Q)} &\;\defeq\;
    \pcont {(A \times B)} {\lam{(a,b)} P\,a \lor Q\,b},
    \\
    \pSum {(\pcont A P)} {(\pcont B Q)} &\;\defeq\;
     \pcont {(A + B)} {[P , Q]}.
  \end{align*}
  It is straightforward to check that products distribute over sums.

  To exhibit weak exponentials, we construct for any two propositional
  containers $\pcont A P$ and $\pcont B Q$ a propositional container $\wExp
  {(\pcont A P)} {(\pcont B Q)}$, abbreviated here as $\pcont E W$, a morphism
  $
    \wev : \pProd {(\pcont E W)} {(\pcont A P)} \to \pcont B Q
  $,
  and a weak currying operation taking an $f : \pProd {(\pcont C R)} {(\pcont A P)}
  \to \pcont B Q$ to $\wcurry{f} : \pcont C R \to \pcont E W$ such that the following diagram
  commutes:
  \begin{equation*}
    \xymatrix@C=2em@R=2em@M=0.5em@L=0.5em{
      {\pProd {(\pcont E W)} {(\pcont A P)}}
      \ar[drr]^{\wev}
      \\
      {\pProd {(\pcont C R)} {(\pcont A P)}}
      \ar[rr]_(0.6){f}
      \ar[u]^{\pProd{\wcurry{f}~}{~\id}}
      & &
      {\pcont B Q}
    }
  \end{equation*}
  Weakness stems from the fact that~$\wcurry{f}$ need not be the unique
  morphism fitting into the above diagram.
  The weak exponential object is defined as
  \begin{multline*}
    \wExp {(\pcont A P)} {(\pcont B Q)} \;\defeq\; \\
    \begin{aligned}[t]
      & \{(k, K) : (A \to B) \times \Prop \mid
         \all{a \of A} Q (k\,a) \to P\,a \lor K
        \} \pcontsymbol {} \\
      & \left(\lam{(k, K)} \some{a \of A} Q (k\,a) \land K\right),
    \end{aligned}
  \end{multline*}
  while evaluation and currying are given by
  \[
  \begin{array}{l c l}
    \wev ((k, K) , a) & \defeq & k\,a, \\[3pt]
    \wcurry{f} \, c & \defeq & ((\lam{a} f (c, a)), R\,c).
  \end{array}
  \]
  Checking that these are morphisms in $\PCont$ and that the above diagram
  commutes is straightforward.
\end{proof}

\begin{corollary}
  The preorder of functional instance reducibilities is a Heyting prealgebra.
\end{corollary}

\begin{proof}
  The preorder reflection converts finite products and sums to finite infima and suprema, respectively,
  and weak exponentials to exponentials, which are just Heyting implications.
\end{proof}

One might hope to improve the weak exponentials in \cref{thm:pcont-distr-wexp}
to proper ones. This is not possible in general, but we may characterise the exponentiable\footnote{An object $X$ is said to be exponentiable when the exponents $Y^X$ exist for all objects $Y$.} propositional containers.
Say that $\pcont A P$ is \emph{decidable}, when
$\all{a \of A} P\, a \lor \lnot P\, a$.

\begin{thm}
  A propositional container is exponentiable if, and only if, it is decidable.
\end{thm}

\begin{proof}
  Let us first observe that an exponential $\pcont E V$ of $\pcont A P$ and $\pcont B Q$, if it exists, takes a certain form.
  Define $\pId \defeq \pcont{\One}{\lam{\_} \top}$ and note the following chain of bijective correspondences:
  \begin{align*}
    E \cong (\pId \to \pcont{E}{V}) \cong (\pProd{\pId}{(\pcont A P)}\to \pcont B Q) \cong (A \to B).
  \end{align*}
  Consequently, without loss of generality we may presume that $E = A \to B$, and also that evaluation
  and currying coincide with the underlying evaluation and currying on~$\Type$.
  Existence of the exponential thus amounts to finding a predicate~$V$ on~$A \to B$ for which
  the usual evaluation is a container map, and the usual currying operates at the level of container maps.

  If $\pcont A P$ is decidable, then for any $\pcont B Q$ the exponential
  $\pExp{(\pcont A P)}{(\pcont B Q)}$ is formed by taking shapes $A \to B$ and the predicate
  $V \, u \defeq \some{a \of A} Q(u\, a) \land \lnot P\, a$.
  It is easy to verify that with this~$V$ currying takes container maps to container maps,
  while decidability of~$P$ is used to establish that evaluation is a container map.

  Conversely, suppose $\pcont A P$ is exponentiable. Given any $a : A$, we decide $P\,a$ as follows.
  Let $\pcont {(A \to A)} V$ be the exponential $\pExp{(\pcont A P)}{(\pcont A {\lam{a'} a' = a})}$.
  The fact that evaluation is a container map yields $V\,\id \lor P\,a$,
  so it suffices to prove $P\,a \lthen \neg V\,\id$, as then $P\,a$ is complemented by $V\,\id$, hence decidable.
  If $P\, a$ holds, then the second projection $\pi_2 : \One \times A \to A$ is a container map $\pProd{\pOne}{(\pcont A P)} \to (\pcont A {\lam{a'} a' = a})$, and so too is its currying $\star \mapsto \id$, which entails $V\,\id \lthen \bot$.
\end{proof}

With the characterization in hand, we may state precisely what it takes for~$\PCont$ to be cartesian closed.

\begin{corollary}
  The category of propositional containers is cartesian closed if, and only if, excluded middle is valid in~$\Prop$.
\end{corollary}

\begin{proof}
  If $\Prop$ validates excluded middle then every propositional container is decidable, hence exponentiable. For the converse, observe that for any $p : \Prop$, exponentiability of $\pcont \One {\lam{\_} p}$ yields decidability of~$p \lor \neg p$.
\end{proof}

Next, we investigate how general instance reducibility~\eqref{eq:instance-reducibility} relates to comodule representations.
To work towards an answer, we write $(\pcont A P) \ileq (\pcont B Q)$ for~\eqref{eq:instance-reducibility},
note that~$\ileq$ is a (large) preorder on~$\PCont$, and refer to~\cite{Bauer:InstanceReducibility} for further details.

Let $\IPow{A} \defeq \{v : A \to \Prop \mid \some{x \of A} v\,x\}$ be the \emph{inhabited powerset} of~$A$.
It is a functor whose action on $f : A \to B$ is given by
\begin{equation*}
  \IPow{f} \, u \defeq \lam{b \of B} (\some{a \of A} u\,a \land f\,a = b).
\end{equation*}
The functor is the carrier of a monad whose unit $\eta_A\,a \defeq \lam{a'\of A} (a = a')$ forms singletons, and that extends a map $f : A \to \IPow{B}$ to $\bind{f} : \IPow{A} \to \IPow{B}$ as
\begin{equation*}
  \bind{f} u \defeq \lam{b \of B} (\some{a \of A} u\,a \land f\,a\,b).
\end{equation*}
Moreover, there is a weak Mendler-style $\IPow$-algebra on $\Prop$ whose extension operation takes a family $P : A \to \Prop$ to
$\palg P : \IPow{A} \to \Prop$, where
\begin{equation*}
  \palg P u \defeq (\some{a \of A} u\,a \land P\,a).
\end{equation*}
The functorial action and the lax algebra laws from \cref{def:weak-mendler-style-T-algebra} amount to the following statements, where $P, Q : A \to \Prop$, $R : B \to \Prop$, and $f : A \to \IPow B$:
\begin{itemize}
\item $(\all{a \of A} P\,a \lthen Q\,a) \lthen (\all{u \of \IPow{A}} \palg P u \lthen \palg Q u)$,
\item $\all{a \of A} \palg P (\eta_A\,a) \lthen P\,a$,
\item $\all{u \of \IPow A} \palg R (\bind f\, u) \to \palg{(\palg R \comp f)} u$.
\end{itemize}
These are intuitionistically valid statements, therefore inhabited. The coherence laws hold automatically because all diagrams in~$\Prop$ commute, by proof irrelevance.
In the following, we write $\IPow^\textrm{c}$ for the resulting monad on $\PCont$.

As expected, there is a right $\IPow^\textrm{c}$-comodule structure on $\cosem{{-}} : \PCont^\op \to \Prop$ with the structure map~$\cook^{+} : \cosem{-} \to \cosem{\IPow^\textrm{c}(-)}$, whose component
\begin{equation}
  \label{eq:cook-instance}%
  \cook^{+}_{\pcont A P} :
  (\all{a \of A} P\,a) \lthen (\all{u \of \IPow\,A} \palg P u)
\end{equation}
at $\pcont A P$ is the (unique) intuitionistic proof inhabiting the above statement. Once again, the comodule laws hold automatically as they take place in~$\Prop$.

Now, a representation of $F : (\all{a \of A} P\,a) \lthen (\all{b \of B} Q\,b)$ with respect to the comodule just described is a map $\tree_F : B \to \IPow{A}$ and a proof
\begin{equation*}
  \eat_F : \all{b \of B}  \palg P {(\tree_F\, b)} \lthen Q\,b,
\end{equation*}
or equivalently, after unfolding the definition of~$\palg P {}$, a proof
\begin{equation*}
  \eat_F : \all{b \of B}  (\some{a \of A} \tree_F\,b\,a \land P\,a) \lthen Q\,b.
\end{equation*}
Observing that~\eqref{eq:instance-reducibility} is intuitionistically equivalent to
\begin{equation*}
  \some{u \of B \to \IPow{A}}
  \all{b \of B}
  (\some{a \of A} u\,b\,a \land P\,a) \lthen Q\,b,
\end{equation*}
we see that once again we have a case of preorder reflection:

\begin{prop}
  The preorder~$\ileq$ of instance reducibilities is the reflection of the
  Kleisli category ${\PCont}_{\IPow{}^\textrm{c}}$ for the inhabited powerset
  monad $\IPow{}^\textrm{c}$ on $\PCont$.
\end{prop}

The structure of ${\PCont}_{\IPow{}^\textrm{c}}$ reflects onto the structure on instance reducibilities.

\begin{corollary}
  The preorder of instance reducibilities forms a Heyting prealgebra.
\end{corollary}

\begin{proof}
  The statement was already established in~\cite[Prop.~2.4]{Bauer:InstanceReducibility}, but it is instructive to observe how the structure arises as the reflection of the corresponding one on the Kleisli category ${\PCont}_{\IPow{}^\textrm{c}}$.
  Since $\IPow^\textrm{c}$ is strong and commutative, finite sums and products in ${\PCont}_{\IPow{}^\textrm{c}}$ are computed
  as in $\PCont$ at the level of objects, and suitably rectified at the level of morphisms~\cite{szigeti83:_kleis,Mulry:LiftingTheorems}. 
  The weak exponential of $\pcont A P$ and $\pcont B Q$ is computed as
  $\pExp {(\pcont A P)} {\IPow{}^\textrm{c} (\pcont {B} {Q})}$.
\end{proof}


\section{Related Work}
\label{sec:related-work}

Trees representing functionals $F : \Nat^\Nat \to \Nat$ are known in the literature as \emph{dialogue
trees}~\cite{escardo13:_contin_system_t_defin_funct_effec_forcin}.
They differ from the older \emph{Brouwer trees}~\cite{troelstra88:_const} in that the latter always ask questions in order $0, 1, 2, \ldots$ Consequently, the nodes need not be explicitly labelled with questions, as we may read those off from the height of the node.
It is unclear how to usefully generalise Brouwer trees to representations of general
functionals $F : \left(\prd{a \of A} P\, a \right) \to \left(\prd{b \of B} Q\, b
\right)$, as one would essentially have to restrict~$A$ to an initial segment of~$\Nat$.

Tree representations of functionals of the form $F : \left(\prd{a \of A} P\, a \right) \to Q$ are the basis of the recent work of Baillon et al.~\cite{baillon22:_garden_pythia_model_contin_depen_settin,baillon23:_contin_type_theor}
that extends to a dependently typed setting the proof that all definable second-order System~T functionals are continuous.
The trees considered therein may use any type of questions and dependently typed answers, like ours, but keep the leaves labelled with answers. Consequently, the connection with containers and Kleisli container morphism is not made.

The generalisation considered by Ghani et al.~\citep{GhaniHP:ContinousFunctions,GhaniHP:StreamProcessors} is of a different kind.
Starting from the observation that $\Nat^\Nat$ is isomorphic to the coinductive type
of streams $\mathsf{Str}\, \Nat \defeq \nu X .\, \Nat \times X$,
they study continuous functionals $F : \nu X . S\, X \to Q$, where $S$ is an endofunctor on $\Set$
expressible as a power-series, i.e., $S\, X \cong \sm{i \of \Nat} A_i \times
X^i$, which amounts to an interpretation of the container $\cont {\sm{i \of
\Nat} A_i} {\lam{(i,a)} \finset{0, 1, \ldots, i}}$. Such functionals are represented by a certain kind of well-founded leaf-labelled trees.
Ghani et al.~also investigate combining such representations to represent and compose stream processors of the form $F :
\mathsf{Str}\, P \to \mathsf{Str}\, Q$.

On the topic of effectful functionals with effect-free tree representations from
\cref{sec:effectful-functionals-with-effect-free-representations}, we mention the work of Jaskelioff
and O'Connor~\cite{JaskelioffO:RepresentationTheorem}, who established a
representation theorem for effectful simply-typed functionals, taking the form
\[
  \int_S (A \to U S P) \to U S Q
  \quad\cong\quad
  U R_{A,P}^* Q. 
\]
Here $R_{A,P}\, X = A \times (P \to X)$ and $(-)^* \dashv U : \D \to \C$, where
$\C$ is small and $\D$ is a full subcategory of $\Set^\Set$. Intuitively, 
the left-hand side corresponds to second-order functionals that are polymorphic in the functor~$S$, and the right-hand side to well-founded trees with $A$-labelled $P$-branching nodes, and $Q$-labelled leaves.

Yet another approach to the study of continuous second-order functionals was provided by
Garner~\cite{Garner:StreamProcessorsAndComodels}, who shows that,
for given sets~$A$ and~$B$, the set of continuous functionals $A^\Nat \to B^\Nat$ is
the underlying set of the terminal $B$-ary comagma in the category of $A$-ary magmas.


\section{Conclusion and Future Work}
\label{sec:conclusion}

Starting from an old observation of Brouwer's about representations of continuous functionals, we have been lead to develop a general theory of representations of second-order functionals in a dependently typed setting, based on a natural notion of a right comodule for a monad on the category of containers.
The examples, generalisations, and variations of the notion that we found suggest that our framework is worthy of further study. We briefly discuss the most promising directions.

Our construction of monads on containers from weak Mendler-style algebras is a useful device, but it only
ever results in a particular kind of a monad that operates on shapes independently from the positions.
Already our prime example, the a monad of trees, is not of this form. It would be useful for several reasons to have a better understanding of general monads on containers, starting with mathematically relevant examples that are not of the weak Mendler-style.

Our treatment of effectful tree-represented and functional functionals must be seen only as a first step towards
a general account of representations of effectful functionals. Apart from improving our understanding of effectful computations, we expect it to have applications in computability theory, as many kinds of computability-theoretic reductions are in essence second-order functionals. One particular kind of computational effect that has eluded us, and is thus hardly addressed in this paper, is partiality. Suffice it to say that naive attempts at obtaining representations of partial functionals in terms of \emph{non}-well-founded trees do not seem to work.

With regards to instance reductions, one can hope that the container-based approach can contribute to the study of instance and  Weihrauch degrees. The formula for the weak exponentials of propositional containers should be looked at closely, as it may help compute interesting examples of Heyting implications between instance degrees, of which at present we know none.

Finally, the practically minded readers, which we consider ourselves to be too, will wonder whether the present work bares any importance on, produces fresh ideas about, or generates new programming techniques for (functional) programming with second-order functionals.



\label{sect:bib}

\bibliographystyle{elsarticle-num} 
\bibliography{references}

\appendix
\section{Definition of the Tree Monad on Containers}
\label{appendix:definition-of-tree-monad}

In this appendix we provide a detailed definition of the \emph{tree monad} on
the category of containers, discussed in
\cref{sec:tree-representation,sec:comodule-representation-of-functionals}. While
the proofs that this definition forms a monad can already be found in
\cite[Theorem~4.5]{GambinoK:PolyMonads}, we include the definition for the 
benefit of the more functional programming minded reader.

We begin by recalling from Section~\ref{sec:comodule-representation-of-functionals} that the carrier of this monad is given by taking $T (\cont A P)
\defeq \cont {\Tree{A,P}} {(\lam{t} \Path[A,P]{t})}$, where the type $\Tree{A,P}$
is given inductively by the following two cases
\[
  \begin{array}{c@{\qquad} c}
    \inferrule 
      {\phantom{\Tree{A,P}}} 
      {\leaf : \Tree{A,P}}
    &
    \inferrule 
      {a : A \quad \trs{t} : P\, a \to \Tree{A,P}} 
      {\node(a,\trs{t}) : \Tree{A,P}}
  \end{array}
\]
and the type family $\PathFam[A,P]$ is given inductively by the following 
two cases
\[
  \begin{array}{c@{\qquad} c}
    \inferrule 
     {\phantom{\Path[A,P]{t}}} 
     {\pstop : \Path[A,P]{\leaf}}
    &
    \inferrule 
     {p : P\, a \quad \pth{p} : \Path[A,P]{\trs{t}\, p}} 
     {\pstep(p,\pth{p}) : \Path[A,P]{\node(a,\trs{t})}}
  \end{array}
\]

For any $\cont A P$, we define the component $\eta_{\cont A
P} : \cont A P \to T(\cont A P)$ of the unit as the container morphism
$\eta_{\cont A P} \defeq \cont {\shapeMapSup{\eta_{\cont A P}}} {\positionMapSup{\eta_{\cont A
P}}}$, where the shape map is given by
\[
  \begin{array}{l@{~~} c@{~~} l}
    \multicolumn{3}{l}{
      \shapeMapSup{\eta_{\cont A P}} ~:~ A \to \Tree{A,P}, 
    }
    \\[5pt]
    \shapeMapSup{\eta_{\cont A P}}\, a & \defeq & \node(a, \lam{\_} \leaf), 
  \end{array}
\]
and the position map is defined as
\[
  \begin{array}{l@{~~} c@{~~} l}
    \multicolumn{3}{l}{
      \positionMapSup{\eta_{\cont A P}} ~:~ \iprd{a \of A} \Path[A,P]{\shapeMapSup{\eta_{\cont A P}}\, a} \to P\, a, 
    }
    \\[5pt]
    \positionMapSup{\eta_{\cont A P}}\, (\pstep(p,\pstop)) & \defeq & p.
  \end{array}
\]

In order to define the Kleisli extension $\bind{(-)}$, we first define
an operation $\graft\, t\, \trs{u}$ that grafts a family of trees $\trs{u} :
\Path[A,P]{t} \to \Tree{A,P}$ into the leaves of a tree $t : \Tree{A,P}$:
\[
  \begin{array}{l@{~~} c@{~~} l}
    \multicolumn{3}{l}{
      \graft_{\cont A P} ~:~ \prd {t \of \Tree{A,P}} (\Path[A,P]{t} \to \Tree{A,P}) \to \Tree{A,P}, 
    }
    \\[3pt]
    \graft_{\cont A P}\, \leaf\, \trs{u} & \defeq & \trs{u}\, \pstop, 
    \\[3pt]
    \graft_{\cont A P}\, (\node(a,\trs{t}))\, \trs{u} & \defeq & 
      \node\big(a, \lam{p \of P\, a} \graft_{\cont A P}\, (\trs{t}\, p)\, 
        (\lam{\pth{p}} \trs{u}\, (\pstep(p, \pth{p}))) \big).
  \end{array}
\]

A path in a grafted tree $\graft_{\cont A P}\, t\, \trs{u}$ comprises two parts: the first one is a path in the base tree~$t$ to one of its leaves, and the second a path in the tree grafted at the leaf where that the first part leads to. We define auxiliary maps that decompose such a path into its constituent parts:
\[
  \begin{array}{l@{~~} c@{~~} l}
    \multicolumn{3}{l}{
      \pfst_{\cont A P} ~:~ 
        \iprd{t \of \Tree{A,P}} \iprd{\trs{u} \of \Path[A,P]{t} \to \Tree{A,P}}, 
    }
    \\[3pt]
    && ~~\Path[A,P]{\graft_{\cont A P}\, t\, \trs{u}} \to \Path[A,P]{t}, 
    \\[3pt]
    \pfst_{\cont A P}\, \ia{\leaf}\, \ia{\trs{u}}\, \pth{p} & \defeq & \pstop, 
    \\[3pt]
    \pfst_{\cont A P}\, \ia{\node(a,\trs{t})}\, \ia{\trs{u}}\, (\pstep(p,\pth{p})) & \defeq &
    \pstep\big(p,\pfst_{\cont A P}\, \ia{\trs{t}\, p}\, \pth{p}\big), 
  \end{array}
\]
and
\[
  \begin{array}{l@{~~} c@{~~} l}
    \multicolumn{3}{l}{
      \psnd_{\cont A P} ~:~ 
        \iprd{t \of \Tree{A,P}} \iprd{\trs{u} \of \Path[A,P]{t} \to \Tree{A,P}} 
          \prd{\pth{p} \of \Path[A,P]{\graft_{\cont A P}\, t\, \trs{u}}} 
    }
    \\[3pt]
    \multicolumn{3}{r}{
      \qquad\qquad\quad \Path[A,P]{\trs{u}\, (\pfst_{\cont A P}\, \pth{p})}, 
    }
    \\[3pt]
    \psnd_{\cont A P}\, \ia{\leaf}\, \ia{\trs{u}}\, \pth{p} & \defeq & \pth{p}, 
    \\[3pt]
    \psnd_{\cont A P}\, \ia{\node(a,\trs{t})}\, \ia{\trs{u}}\, (\pstep(p,\pth{p})) & \defeq &
    \psnd_{\cont A P}\, \ia{\trs{t}\, p}\, \ia{\lam{\pth{q}} \trs{u}\, (\pstep(p, \pth{q}))}\, \pth{p}.
  \end{array}
\]
The idea is that a path in a tree $\graft_{\cont A P}\, t\, \trs{u}$ consists of
a path in $t$ composed with a path in one of the trees $\trs{u}$, and $\pfst$
and $\psnd$ then respectively return these paths.

Using these operations, we define the
Kleisli extension $\bind{(\cont f g)} : T(\cont A P) \to T(\cont B Q)$ of
$\cont f g : \cont A P \to T(\cont B Q)$ to be
$\bind{(\cont f g)} \defeq \cont {\shapeMap{\bind{(\cont f g)}}} {\positionMap{\bind{(\cont f g)}}}$, where the shape map is defined as
\[
  \begin{array}{l@{~~} c@{~~} l}
    \multicolumn{3}{l}{
      \shapeMap{\bind{(\cont f g)}} : \Tree{A,P} \to \Tree{B,Q}, 
    }
    \\[5pt]
    \shapeMap{\bind{(\cont f g)}}\, \leaf & \defeq & \leaf, 
    \\[5pt]
    \shapeMap{\bind{(\cont f g)}}\, (\node(a,\trs{t})) & \defeq & 
    \\
    \multicolumn{3}{l}{
      \qquad\qquad\qquad\quad
      \graft_{\cont B Q}\, (f\, a)\, (\lam{\pth{q} : \Path[B,Q]{f\, a}} \shapeMap{\bind{(\cont f g)}}\, (\trs{t}\, (g\, \ia{a}\, \pth{q}))), 
    }
  \end{array}
\]
and the position map is defined as
\[
  \begin{array}{l@{~~} c@{~~} l}
    \multicolumn{3}{l}{
      \positionMap{\bind{(\cont f g)}} : \iprd{t \of \Tree{A,P}} \Path[B,Q]{\shapeMap{\bind{(\cont f g)}}\, t} \to \Path[A,P]{t}, 
    }
    \\[5pt]
    \positionMap{\bind{(\cont f g)}}\, \ia{\leaf}\, \pth{q} & \defeq & \pstop, 
    \\[5pt]
    \positionMap{\bind{(\cont f g)}}\, \ia{\node(a,\trs{t})}\, \pth{q} & \defeq & 
      \pstep\big(g\, \ia{a}\, (\pfst_{\cont B Q}\, \pth{q}), 
      \positionMap{\bind{(\cont f g)}}\, (\psnd_{\cont B Q}\, \pth{q}) \big).
  \end{array}
\]
Intuitively, the shape map $\shapeMap{\bind{(\cont f g)}}$ replaces every node
labelled with $a : A$ in the given $(A,P)$-tree by the $(B,Q)$-tree $f\, a$,
after which it uses the $\graft$ operation to graft recursively computed
$(B,Q)$-subtrees into the leaves of $f\, a$. The position map
$\positionMap{\bind{(\cont f g)}}$ uses the projection operations and $g$ to
turn paths in $(B,Q)$-trees of the form $f\, a$ into single steps in the given
$(A,P)$-tree at nodes labelled with $a$.


\end{document}